\documentclass[useAMS,usenatbib,usegraphicx]{mn2e}

\title[High-resolution stellar population models]{Stellar population models at high spectral resolution}
\author[C. Maraston and G. Str\"omb\"ack]{C. Maraston$^{1}$\thanks{E-mail:
claudia.maraston@port.ac.uk; models available at www.maraston.eu or at request 
from author} and G. Str\"omb\"ack$^{1}$\\
$^{1}$University of Portsmouth, Dennis Sciama Building, Burnaby 
Road, Portsmouth PO1 3FX\\}
\begin{document}
\bibliographystyle{mn2e}
\date{Accepted 2011 August 30. Received 2011 August 30; in original form 2010 October 15}
\pagerange{\pageref{firstpage}--\pageref{lastpage}} \pubyear{}
\maketitle
\label{firstpage}
\begin{abstract}
We present new, high-to-intermediate spectral resolution stellar population models, based on four popular libraries of empirical stellar spectra, namely Pickles, ELODIE, STELIB and MILES. These new models are the same as our previous models, but with higher resolution and based on empirical stellar spectra, while keeping other ingredients the same including the stellar energetics, the atmospheric parameters and the treatment of the Thermally-Pulsating Asymptotic Giant Branch and the Horizontal Branch morphology. We further compute very high resolution (R=20,000) models based on the theoretical stellar library MARCS which extends to the near-infrared. 
We therefore provide {\it merged} high resolution stellar population models, extending from $\sim~1000$~\AA\ to 25,000~\AA, using our previously published high resolution theoretical models which extended to the ultraviolet.
We compare how these libraries perform in stellar population models and highlight spectral regions where discrepancies are found. We confirm our previous findings that the flux around the $V$ band is lower (in a normalised sense) in models based on empirical libraries than in those based on the BaSeL-Kurucz library, which results in a bluer $B-V$ colour. Most noticeably the theoretical library MARCS gives results fully consistent with the empirical libraries. This same effect is also found in other models using MILES, namely Vazdekis et al. and Conroy \& Gunn, even though the latter authors reach the opposite conclusion. The bluer predicted $B-V$ colour (by 0.05 magnitudes in our models) is in better agreement with both the colours of Luminous Red Galaxies and globular cluster data. We test the models on their ability to reproduce, through full spectral fitting, the ages and metallicities of galactic globular clusters as derived from CMD fitting and find overall good agreement. We also discuss extensively the Lick indices calculated directly on the integrated MILES-based SEDs and compare them with element ratio sensitive index models. We find a good agreement between the two models, if the metallicity dependent chemical pattern of the Milky Way stars is taken into account in this comparison. As a consequence, the ages and metallicities of galactic globular clusters are not well reproduced when one uses straight the MILES-based indices, because subtle chemical effects on individual lines dominate the age derivation. The best agreement with the ages of the calibrating GCs is found with either element-ratio sensitive absorption-line models or with the full SED fitting, for which no particular weight is given to selected lines.
\end{abstract}
\section{Introduction}
Galaxy evolution studies are reaching a high level of sophistication due to the very high quality of observational data permitted by modern technology, and the level of spectral details that such observations carry in (just to name a few examples, the Sloan Digital Sky Survey - SDSS - Eisenstein al. 2011, Aihara et al. 2011; the GAlaxy Mass Assembly - GAMA - Driver et al. 2011; RESOLVE, Kannappan et al. 2011; GMASS, Cimatti et al. 2008; the SINS survey, 
F{\"o}rster-Schreiber et al. 2011)\nocite{eisetal11,dr8,drietal11,kanetal11,cimetal08,sins}. In this era of {\it precision astrophysics}, interpretative models, such as stellar population and galaxy models, need to keep pace with the fast observational development. 

Stellar population models allow the calculation of the integrated spectral energy distribution (SED) of a synthetic stellar population - a star cluster or a galaxy - for arbitrary choices of parameters such as the age, the Initial Mass Function (IMF), the metallicity, the star formation history, the Horizontal Branch morphology, etc. Models are based on three ingredients, namely the stellar energetics, the stellar spectra - onto which the energetics are distributed -  and the computational method adopted for calculating the integrated model.  
At given energetics, a model SED of higher spectral resolution is obtained by employing higher resolution stellar spectra. In this paper we present the higher-resolution version of the Maraston (2005) stellar population models.

In the Maraston (2005) models, the library of low-resolution, theoretical stellar spectra compiled by \citet{lejetal97,lejetal98}, was adopted. This 
is largely based on the classical model atmospheres of \citet{kur79} and a patchwork of different calculations for cool, M-type giants and dwarfs. The final assembly was further corrected in order to match empirical relations between stellar temperature and colours. The library does not contain Thermally-Pulsating Asymptotic Giant Branch (TP-AGB) carbon-rich and oxygen-rich stars, for which in Maraston (2005) the empirical, time-averaged spectra of C- and O-type stars from \citet{lanmou02} were utilised.

A higher spectral resolution is required for a detailed modelling of emission and absorption lines. This is not the only necessary feature because element abundance ratios play an equally important role, but it is the first obvious step to resolve the lines properly. In turn, a high spectral resolution SED can be obtained either with theoretical or empirical stellar spectra. The two approaches have both pros and cons. While theoretical spectra give in principle high flexibility on the stellar parameter coverage (temperature, gravity, metallicity) and in the wavelength extension, known problems exist in the modelling of specific lines \citep[e.g. the $H_{\beta}$~line at $\sim~4861$~\AA,][]{coeetal07, waletal09} as well as continuum regions due to line blanketing \citep[around the $V$~band,][]{wor94} and around the $g,r$~SDSS bands \citep[][see below]{CMetal09a}. Empirical stellar spectra have all lines that are observed in stars, but their intensity and flux ratios depend on the chemical enrichment history of the site where the star is found. Moreover, empirical libraries usually cover a limited region of stellar parameter space and a limited $\lambda$~range. Hence, the approach of combining empirical and theoretical spectra is the most convenient one. 

The first version of high-resolution Maraston-type models was presented in \citet{CMetal09b}, where fully theoretical, very high-resolution SEDs in the mid- and far-$UV$~($\lambda < 4350$) were published. At optical wavelengths, several empirical stellar libraries have been made available in the recent years, containing enough stars for nearly-complete stellar population models in a relatively broad wavelength range to be calculated (the Pickles library, Pickles 1998; the Jones's library, Jones 1999; the STELIB library, Le Borgne et al. 2003; the ELODIE library Prugniel \& Soubiran 2001, and the MILES library, Sanchez-Blazques et al. 2006)\nocite{pic98,jon99,leboretal03,prusou01,patriciaetal06}. Stellar population models based on these libraries exist in the literature (e.g. Bruzual \& Charlot 2003 for the STELIB library; Vazdekis et al. 2010 for the MILES library).\nocite{brucha03,vazetal10}

Our preliminary attempt with these libraries \citep{CMetal09a} allowed us to improve upon a long-standing mismatch between the colours of {\it luminous red galaxies} in SDSS and stellar population models based on theoretical stellar spectra \citep{eisetal01,waketal06}. Also, a few remaining problems with the broadband colours of stellar population models, such as the ({\it B-V}) colours of Milky Way GCs \citep[e.g.][]{wor94,CM05} could be connected to the same origin, and we shall show in this article that this was indeed the case. 

In this paper we present the full grid of M05-type models for a number of these different optical empirical libraries. In addition, we also employ the theoretical MARCS library for modelling the spectra of old populations with a high resolution up to the near-IR. We shall test the models with GCs data, and compare them with results from other authors.

The paper is organised as follows. In Section~\ref{eps} we recall the main features of the evolutionary population synthesis code and in Section~\ref{libraries} those of the empirical spectral libraries. In Section~\ref{scaling} we describe the implementation of the empirical stellar spectra in the synthesis code, and also the calculation of models based on high-resolution theoretical spectra. Stellar population models are presented and amply discussed in Section~\ref{sec:ssp} including a comparison with the literature, while in Section~\ref{sec:testing} the model fit to globular cluster data is treated. A summary is given in Section~\ref{summary}.
\section{Summary of the population synthesis code}
\label{eps}
The foundation upon which these models have been built was 
thoroughly described in \citet[hereafter M98 and M05, respectively]{CM98,CM05}. 
Here, we present a brief recapitulation of those models.

Evolutionary population synthesis models assume a stellar evolution prescription, which in this case consisted of the isochrones and stellar tracks by
 \citet{casetal97} for ages larger than $\sim~30$~Myr  and by \citet{schetal92} for younger populations. Sets of models have been also computed with the Padova stellar evolutionary models \citep{giretal00}. The theoretical fuel for the TP-AGB phase - from \citet{ren92} -  was calibrated with observational data. Reimers-type mass-loss was applied to the Red Giant Branch track in order to generate blue and intermediate Horizontal Branch (HB) morphologies as observed in Milky Way globular clusters, and also blue HBs at high metallicity.
 
The synthesis technique exploits the fuel consumption approach~\citep{ren81} for calculating the energetics of post Main Sequence phases, i.e. the amount of fuel available for nuclear burning using the evolutionary track of the turnoff mass. In the M05 models, the fuel is used as integration variable in post Main Sequence. For the Main Sequence, the standard isochrone synthesis technique is applied, where the integration variable is the mass. 

As for the stellar spectra, the BaSeL assembly \citep{lejetal97} of the \citet{kur79} library was employed for all evolutionary phases except the TP-AGB, for which empirical carbon-rich and oxygen-rich stellar spectra were used \citep{lanmou02}. 
Finally, the code structure is modular (see Maraston 1998), i.e. the three ingredients -
the energetics, the atmospheric parameters temperature and gravity, and the individual stellar spectral - are stored in independent matrices, which allows one to easily experiment on their differential impact on the final models. This structure was particularly useful to compute the models of this paper for which we only vary the spectral transformation matrix.
\section{The libraries of empirical stellar spectra}
\label{libraries}
Four different libraries of flux-calibrated empirical stellar spectra have been considered, namely \citet{pic98}, ELODIE, STELIB, and MILES.
A brief summary of the properties of these libraries can be found in 
Table \ref{tab:libs}. Note that the values of T$_{eff}$, log(g), and [Fe/H] 
reported there are minimum and maximum values only, and therefore say little about the actual sampling of the parameter space. For 
example, in the MILES library only three stars have a metallicity higher than 
[Fe/H]$=+1.0$. The actual range interesting for population synthesis modelling is illustrated in Figures \ref{fig:covpic}-\ref{fig:covelo}.

In Table \ref{tab:libs} we provide two values for the spectral resolution, the {\it nominal} one cited in the reference papers and the {\it actual} one as was determined by later works. For MILES  
we found that the actual resolution of the library is somewhat coarser than the 2.3 \AA\ stated in \citet{patriciaetal06} and \citet{vazetal10} and is instead 2.54~\AA\ \citep[][]{beietal11}. This result has been independently confirmed by the MILES group \citep{falbaretal11}. For STELIB, \citet{macartetal09} found that the resolution of the \citet{brucha03} models is coarser than the resolution of individual STELIB spectra, which is the library adopted in those models (see Section 3.2). 
The actual resolutions should be used.

Below follows a review of important features of each library, together with a description of 
any alterations we may have imposed upon the spectra.

Note that stars from the galactic Bulge are not included in these libraries, and also not spectra of carbon- and/or oxygen-rich TP-AGB stars. For the latter, we use the same empirical spectra as in the M05 models (see Section 3.6). 

\begin{table*}
  \centering
  \begin{minipage}{170mm}
    \caption{Properties of the empirical libraries}
    \label{tab:libs}
    \begin{tabular}{@{}llllllll}
      \hline
      Reference & No. of & Wavelength & nom. FWHM & actual FWHM & T$_{eff}$ (K) & log(g) & [Fe/H]\\
       & stars & range ({\AA}) & (at 5500 {\AA}) & (at 5500 {\AA}) & & & \\
      \hline
      Pickles (1998) & 131 & 1150-25000 & 11 {\AA} & 11 {\AA} & 2500-39800 & 
      [$-$0.5, $+$5.5]\footnote{Inferred from matching to the isochrones of \citet{giretal00} in a {\it M$_{V}$}-{\it (B-V)} CMD. } & [$-$0.8, $+$0.6 ]\\
      STELIB & 249 & 3200-9300 & 3.0 {\AA} & 3.4 {\AA} & 3250-51000 & [0.0, $+$5.0] &
      [$-$2.75, $+$0.6]\\
      MILES & 985 & 3500-7430 & 2.3 {\AA} & 2.54 {\AA} &  2750-36000 & [0.0, $+$5.5] &
      [$-$2.9, $+$1.6]\\
      ELODIE v3.1 & 1388 & 3900-6800 & 0.55 {\AA} & 0.55 {\AA} & 3100-51000 & 
      [$-$0.25, $+$4.9] & [$-$3.0, $+$1.0]\\
      \hline
    \end{tabular}
  \end{minipage}
\end{table*}
\subsection{The Pickles library}
The Pickles library \citep{pic98} differs from the other three in that it 
does not contain spectra of individual stars. Rather, the author has vacuumed 
the literature for a versatile collection of stellar libraries and combined 
the various observations, in order to provide standard spectra for all normal 
spectral types and luminosity classes. Because they are (for the most part) 
averages of spectra from several sources, the Pickles spectra should not be as 
affected by systematic and random observational errors, as a library 
containing single observations of each star using the same setup.

\begin{figure}
  \includegraphics[width=84mm]{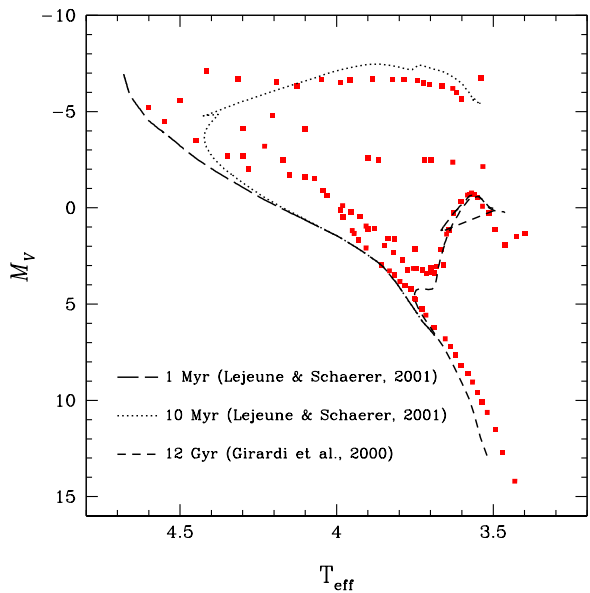}
  \caption{Evolutionary coverage at Z$_{\odot}$ for the Pickles library. 
    Superimposed are the 1 Myr, 10 Myr and 12 Gyr isochrones from \citet{lejsch01} and 
    \citet{giretal00}, respectively.}
  \label{fig:evocov}
\end{figure}

The library has exemplary coverage of evolutionary phases, see Fig.\ref{fig:evocov}, 
from the coolest M dwarfs and giants 
(T$_{eff}\approx$~2500 K) to hot O-type stars and supergiants. Only 
solar metallicity models can be computed from this library, however, since there are 
not enough spectra to cover all evolutionary phases to a satisfactory 
degree for non-solar metallicities. The spectral resolution is R$=500$ and the wavelength range is very wide, ranging from the far-UV (1150~{\AA}) to the K-band (2.5 $\umu$m). The extension into the UV in the Pickles standard 
spectra was accomplished using the low-dispersion {\it International Ultraviolet 
Explorer} ({\it IUE}) spectra \citep{hecetal84}.
Above $\sim$1$\,\umu$m, about half of the spectra lack spectroscopic 
observations, and the author constructed a smooth energy distribution 
from broadband photometry here. This may imply that near-IR absorption features may not be well resolved. 
Also note that, at these wavelengths, the M~giant spectra are synthetic. 

The effective temperatures of the Pickles spectra were assigned by means of 
({\it V-I}$_{C}$) and ({\it V-K}) colour-temperature relations mainly from 
\citet{bes79}, and from \citet{fluetal94} for the M giants. No accurate 
metallicity calibration has been performed, other than sorting the spectra 
into solar or non-solar metallicity based on a recipe described in 
\citet{picvanderkru90}. 
\subsection{STELIB}
\citet{leboretal03} obtained roughly 250 stellar spectra of fairly high resolution 
($\sim3\,${\AA} FWHM) and with good wavelength coverage. Most spectral types 
and luminosity classes are included, with a fair coverage in metallicity. There is a deficit of stars with [Fe/H] $<\,-1.0$. 
Several spectra lack atmospheric parameter determinations, and many do not 
have complete spectroscopic observations over the entire quoted wavelength 
range. In addition to this, all spectra available to the public suffer from 
telluric contamination. We 
have circumvented this problem by simply interpolating over the most affected 
wavelength regions, more precisely 6850-6950 {\AA}, 7580-7700 {\AA}, and 
8850-9050 {\AA}. Granted, this might not be the most aesthetically pleasing 
solution, and it is by no means a complete eradication of all telluric 
absorption features, but it acts as a warning beacon to users and ensures that 
no science can be done within said regions. The STELIB library has previously 
been adopted for stellar population synthesis by \citet[hereafter BC03]{brucha03}, to which we refer for more information. 
These authors replaced the contaminated segments with theoretical Kurucz spectra for stars cooler than 7000 K, and 
with Pickles spectra for hotter stars.

Upon inspection of the individual spectra of the STELIB library, \citet{macartetal09} 
found an rms spread of 1.09 {\AA} in starting wavelength and 0.00033 {\AA} in 
wavelength dispersion. If not corrected for, this will cause the coaddition of 
any spectra into the SED of a stellar population to have a degraded resolution 
compared to the resolution of the individual stellar spectra, acting 
effectively as a velocity broadening. In addition to this, they also found an 
rms error of 11.4 km s$^{-1}$ for the radial velocities adopted by 
\citet{leboretal03}, and small systematic deviations in the wavelength scale, of no 
more than $\pm$0.5 {\AA}. Both of these attributes will further add to the 
downgrading of the resolution of the stellar population model.

In order to rectify these problems with the STELIB wavelengths, thus avoiding 
loosing valuable resolution, we have applied small, non-linear shifts to the 
wavelength scale, derived for each individual STELIB spectrum, by fixing the 
positions of eight strong absorption features, and use interpolation to obtain the 
wavelength shifts for all flux points in between. The spectra were then 
rebinned to our adopted scale of 3201-9296.5 {\AA} in steps of 0.5 {\AA}. The 
reason for straying from the 1 {\AA} step-size is that we found that the 
rebinning was accompanied by a small, but non-negligible, smoothing of some 
absorption features, which could be made insignificant with a finer sampling.

The majority of atmospheric parameters in STELIB were taken from the \citet{cayetal97,cayetal01} 
catalogues of [Fe/H] determinations, which are compiled from a vast number of 
different sources  in the literature. These values have been complemented with 
additional T$_{eff}$ determinations from \citet{blalyn98}, \citet{diben98} and 
\citet{Aloetal96,Aloetal99}. Colour-temperature relations from the latter have also 
been used where available. Several stars, that are also part of the ELODIE 
library \citep{prusou01}, were given the atmospheric parameters listed therein, 
if no other values could be found. If more than one determination of a 
parameter for the same star was available, these were averaged, with a larger 
weight given to the most recent determinations. 
\subsection{MILES}
The MILES library \citep{patriciaetal06} provides spectra for almost 1000 stars, 
covers most evolutionary stages -- based upon what can be expected to exist in 
our Galaxy -- at all metallicities down to [Fe/H] $\sim\,-2.0$. Only stars at 
the extreme ends of the temperature scale are lacking, such as giants cooler 
than around 3400 K, and young stars hotter than $\sim$26000 K. The resolution is 
high (2.54~{\AA} FWHM), and the wavelength range is primarily confined to the 
optical (3500-7400~{\AA}). Values for all atmospheric parameters are taken 
from a large number of sources in the literature, but a careful attempt has 
been made to homogenise the different scales. The reader is referred to 
\citet{cenetal07} for a detailed description of this procedure. When it comes to 
the effective temperatures, however, it should be noted that the reference 
system adopted for this calibration comprises only values within the range
4000-6300 K \citep{cenetal07}. This may imply that the temperatures of stars outside this approximate interval 
may not be as homogeneous, even if an effort is made at tying every temperature to the system defined in the above interval. 

Linking to the discussion in the Introduction, \citet{miletal10} published [Mg/Fe] abundance ratios for roughly 75 \% of the 
stars in the MILES library. The characteristic 'knee' (see their Figure~1), which signals the onset 
of enrichment from SN Ia, occurs at [Fe/H]$\sim\,-$0.9, below which the 
[Mg/Fe]-enhancement lies fairly constant around $+$0.4, albeit with a large 
scatter. Stellar population models incorporating MILES spectra will
therefore follow a similar abundance pattern. The same will happen with the other libraries in the metal-poor regime.
\subsection{ELODIE}
Compared to the original library of ELODIE echelle spectra \citep{prusou01}, the 
latest version, ELODIE v3.1 \citep{pruetal07}, contains roughly twice as many 
stars and a slightly increased wavelength range. The latter is still rather 
narrow, spanning only around 3000 {\AA} in the optical, which is not 
surprising considering the very high resolution (R$\,\approx$10,000). One may 
argue about the relevance of applying stellar population models of such high 
resolution to velocity-broadened galaxy spectra, and for this purpose 
alternative versions of the library with lower resolutions may be computed. 
There are, however, stellar populations in which velocity broadening is less prominent, 
such as dwarf galaxies and star clusters, that might benefit from such a high-resolution model.

Most stars in the ELODIE library have solar or near-solar metallicity. Only 68 
stars out of 1388 are listed as having [Fe/H]$<-1.0$, which limits our ability 
to produce ELODIE-based models at low metallicity. In the original paper 
\citep{prusou01} atmospheric parameters were derived using the TGMET method 
\citep{katetal98}, or taken from the literature 
\citep{caretal94,Aloetal96,Aloetal99,cayetal97,cayetal01,the98,blalyn98}. Colour-temperature 
relations from \citet{Aloetal96,Aloetal99} were also used. Final parameter values 
were taken as averages of multiple determinations, giving lower weight to old 
references. In the latest version \citep{pruetal07}, these parameters have been 
updated using more recent publications.

The stellar parameter determinations and flux calibration of the stars in the ELODIE library have also been assigned quality flags by the authors. We take advantage of this to create as reliable SSP models as possible, by allowing only stars with the highest quality of flux calibration to contribute, and in cases where two or more stars occupy the same location in stellar parameter space, choosing the one with the most reliable parameter estimations. Some stars have also been observed multiple times. Again, in such cases we chose the spectrum with the highest values of the quality flags. If there are two or more spectra of the same star and with the same quality flags we compute an average spectrum for this star, taking also the average of the stellar parameters.
\subsection{The flux calibration}\label{sec:fluxcal}
\begin{figure*}
  \centering
  \begin{minipage}{150mm}
    \includegraphics[width=140mm]{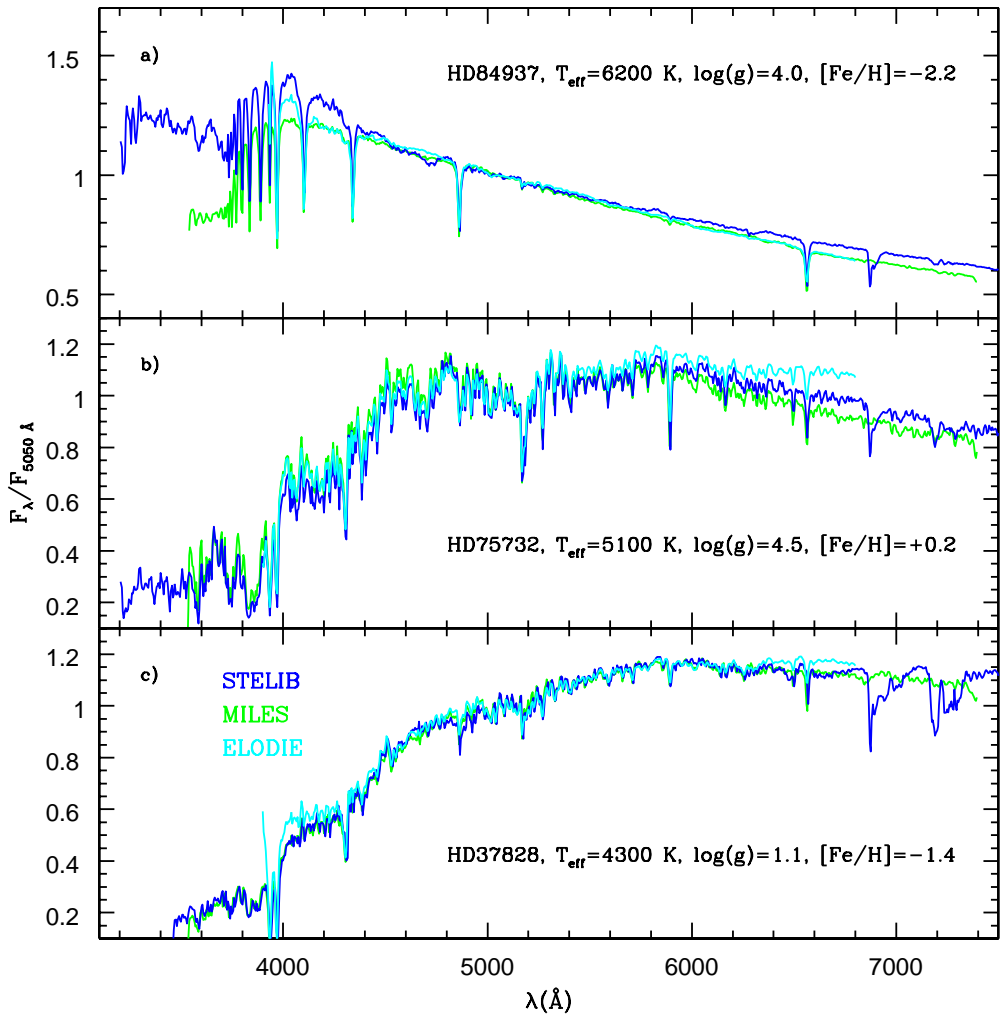}
    \caption{Comparison of the STELIB, 
      MILES, and ELODIE libraries. The three panels depict the spectra of three 
      stars present in all libraries, smoothed to a common, low resolution of 
      R$=$500, where {\it a)} is a metal-poor sub-dwarf; {\it b)} a super-solar 
      metallicity MS star; and {\it c)} a metal-poor giant}
    \label{fig:starcomp}
  \end{minipage}
\end{figure*}

The STELIB, MILES, and ELODIE libraries have a few stars in 
common, which gives us the possibility to obtain some indication of 
the quality of flux calibration. In Fig.\ref{fig:starcomp} we compare 
three stars of intermediate temperatures, chosen to represent a large range in 
metallicity and gravity. Spectra 
have been smoothed to a common, low resolution of R$=$500, and normalized to 
unity at 5050 {\AA}, in order to make any differences in the continuum slope 
more clearly distinguishable. None of the stars are known to be a variable. As 
can be seen in the figure, for at least two of the stars there are 
surprisingly large discrepancies, hinting at the uncertainties involved in 
accurately flux calibrating a stellar spectrum. If the differences in flux 
calibration are due to systematics, these will also be transferred over to the 
SEDs of the stellar population models. Clearly visible are 
the telluric absorption features in the STELIB spectra around 6900 {\AA}.

A careful description of the complex flux calibration process for the ELODIE 
spectra is given by \citet{prusou01}. Since this library is a compilation of 
archive spectra rather than being part of a specific observation programme, no 
standard stars have been observed. The instrumental response is therefore 
evaluated by comparisons to several other low- and high-resolution libraries, 
and the spectra are corrected thereafter. At 3900 {\AA} -- the blue limit of 
the ELODIE spectrograph -- the global efficiency of the instrument drops by a 
factor of 50, which may compromise an accurate flux calibration in the blue 
end of the spectra.

The flux calibration of the STELIB stars was tested in \citet{leboretal03} by 
comparing synthetic photometry against published photometry in the Lausanne 
database \citep{meretal97}. The {\it (B-V)} colour is by far the one with the 
smallest rms dispersion, 0.083 mag, with no discernable offset. The 
{\it (U-B)} and {\it (R-I)} have the largest dispersions, which is due to 
inhomogeneities in the photometric systems, and the fact that the STELIB stars 
do not have the required wavelength coverage for the {\it U}- and 
{\it I}-band filters. There appears to be a slight trend in the {\it (V-R)} 
colour, such that the reddest (coolest) stars appear somewhat bluer in the 
STELIB library. It is argued by the authors, that this is most likely due to 
differences in the photometric systems. It should also be noted that these 
tests were performed on stars not corrected for interstellar reddening. All 
this taken together makes it difficult to assess the accuracy of the flux 
calibration in bands other than {\it B} and {\it V}.

The MILES stars have gone through a rigorous spectrophotometric quality 
control \citep{patriciaetal06}. An internal consistency check, comparing several 
observations of the same stars, revealed that the random rms error affecting 
the flux calibration is around 0.013 mag for the {\it (B-V)} colour. An 
external comparison with the Lausanne photometric database \citep{meretal97}, 
showed a mean offset of 0.015 mag for {\it (B-V)}, in the sense that the MILES 
spectra are bluer. The rms dispersion of 0.025 mag is considerably smaller 
than the one obtained in \citet{leboretal03} for the STELIB stars. For stars in 
common with the STELIB library, the MILES {\it (B-V)} colours are, on average, 
0.01 mag bluer, but with a trend that essentially all stars with 
T$_{eff}<$5600 K are bluer, whereas almost all stars hotter than $\sim$7200 K 
appear redder (see their Fig.11). The authors also defined seven box filters 
with effective wavelengths between 4000 and 6200~{\AA}, in order to compare 
synthetic colours for stars in common with the STELIB and ELODIE libraries. 
For all possible combinations of filters allowed by the wavelength range in 
common between the libraries, any statistically significant offsets show the 
MILES stars to exhibit bluer colours. For colours involving the filters with 
the largest central wavelengths, there once again appears to be a trend of the 
coolest stars being significantly bluer in the MILES library (their Fig.9), 
whereas the situation is the opposite for the hottest stars. It is difficult to 
assess the flux calibration red-wards of around 6500 {\AA}, since this portion 
of the spectra is more or less left out of the quality check.

The trends summarised above will be mirrored in the SSP models (cf. Section~5).
\subsection{On the temperature scales of empirical libraries}\label{sec:tempscale}
In this work we opt to retain the stellar parameters as they have been published by the authors of the various
libraries, because our aim is to assess the joint effect of a library/temperature scale/stellar parameters on stellar population models, with the ultimate goal at understanding different results that are obtained on galaxies using different models. As discussed, the temperature scales are somewhat inhomogeneous, which may cause differences in the integrated stellar population spectra (e.g. Percival \& Salaris 2009) that can sometimes be hard to explain. For example, we could attribute to the adopted temperature scale for red giants part of the reason why the near-IR flux of MILES stars drop longward 6000~\AA~(see Appendix A). 
On the other hand, a dedicated publication is required in order to homogeneise the different temperature scales. Star numbers vary a lot between different libraries, hence a common scale built avoiding extrapolation could be worked out only for a subsample of stars, with the effect of reducing the stars available for population synthesis.
\begin{figure}
  \centering
    \includegraphics[width=0.49\textwidth]{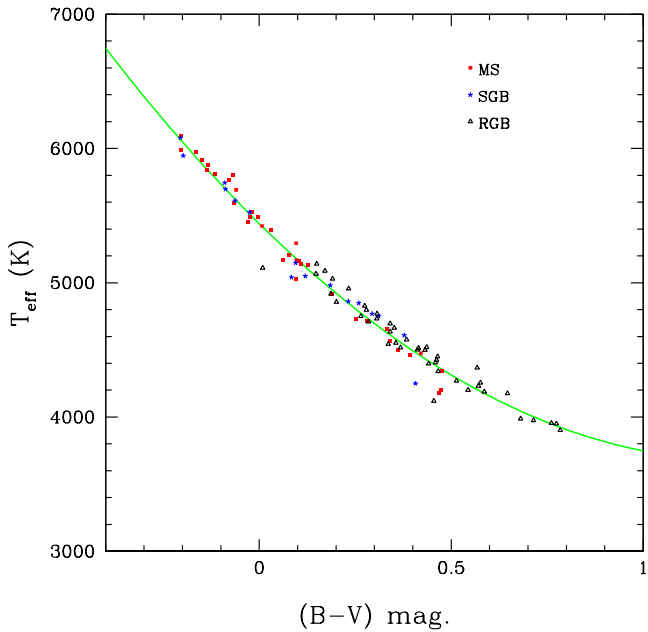}
    \caption{Fit to the $B-V$~colour of MILES stars vs effective temperature.}
    \label{fig:tescalestelibonmil}
\end{figure}

In order to get a sense for the effect of different temperature scales, we used MILES, that contains more stars than the other libraries, and placed the STELIB library on the MILES temperature scale. This was achieved by measuring the $B-V$ colours of the MILES stars at a given metallicity and fit a relation between $B-V$ and temperature. A MILES-based temperature scale for STELIB was then obtained by deriving the temperature of STELIB stars using their $B-V$. The resulting relation is shown in Figure~\ref{fig:tescalestelibonmil}. The effect on the stellar population model will be shown in Section~\ref{sec:oldssp}.
\subsection{Keeping the TP-AGB empirical spectra from M05}
The TP-AGB empirical stellar spectra by \citet{lanmou02} used in the M05 code have a coarser resolution than most of the adopted libraries ($R\sim1100$). For keeping the TP-AGB phase in the new models, we have 
re-binned the TP-AGB spectra to the sampling proper of the various empirical libraries. Except for the Pickles-based models, this implies that at the ages relevant to the TP-AGB phase in the M05 models (0.2 to 2 Gyr) and at $\lambda~>6000$~the actual resolution of the models will be slightly coarser. Note that - because we have not downgraded the TP-AGB spectra - there is an overall improvement in the definition of the carbon and oxygen-rich star spectral breaks in the integrated model compared to M05.
\section{Implementation of the empirical spectra}
\label{scaling}
One of the main advantages of 
using theoretical spectra is that they can be computed to sample the stellar 
parameter space arbitrarily well. Inserting them into a stellar population 
synthesis code is therefore relatively straightforward. When dealing with the 
more sparsely sampled grid of stellar spectra that are found in the empirical 
libraries, however, one has to use a slightly different approach.

\subsection{Assigning empirical stellar spectra to the stellar track}

In the M05 code, a stellar spectra is assigned to each [T$_{eff}$, log(g)]-bin 
 by means of a quadratic interpolation in T$_{eff}$ and log(g) at given metallicity. This approach is generally 
not feasible when empirical libraries are used, due to their coarser sampling 
of the parameter space.

Instead, for each empirical library we created different tables of stellar 
spectra, corresponding to the different evolutionary phases, i.e. MS, SGB, RGB, HB, 
E-AGB, and supergiants. The categorisation was 
performed based mainly on the surface gravities and luminosity classes of the 
spectra, but other discriminants, such as visual inspection of the spectra, 
was used in uncertain cases. More specifically, the cuts in surface gravity 
that were applied for each respective evolutionary phase are: 
MS (log(g)$>$4.0), SGB (3.5$<$log(g)$<$4.0), RGB, HB and E-AGB 
(1.0$<$log(g)$<$3.5), and supergiants (log(g)$<$1.0). These cuts are, of 
course, only approximate and depend somewhat on temperature as well; it is 
not unusual for hot MS stars (spectral class A or B) to have surface gravities 
below 4.0 dex. Spectra on the borderline between two evolutionary phases were 
therefore permitted to be present also in both corresponding tables.

Within one evolutionary phase, the actual 
representative spectrum for each mass bin with its given 
temperature is then calculated by means of linear interpolation of the 
logarithmic fluxes in log(T$_{eff}$). The number of stars assigned to each phase is never smaller  than 10. We have checked the stability of the procedure by using fewer or more stars and fixed the procedure when the results were found to be stable. Note that the whole procedure is also checked at the bottom of the calculations by comparing the integrated models with other calculations based on theoretical stellar spectra or with literature calculations (see Section~\ref{sec:ssp}).

\subsection{Scaling the empirical stellar spectra}
Another difference between theoretical and empirical stellar spectra is 
the flux units. Whereas theoretical spectra preferentially are given in 
absolute units, such as luminosity (ergs s$^{-1}$ {\AA}$^{-1}$) -- or any 
other unit that can easily be converted into luminosity; the main point being 
it makes the population synthesis very straightforward -- things are more 
complicated with empirical spectra. Either fluxes are given as they were 
measured (ergs s$^{-1}$ cm$^{-1}$ \AA$^{-1}$) and thus dependent on the 
distance to the stars, as is the case with STELIB and ELODIE spectra, or they 
have been normalised to unity at some wavelength, a method adopted in the 
Pickles and MILES libraries. In both cases, the true relative energy scale 
between the spectra is washed out.

In order to be able to use a consistent approach for all the 
empirical libraries, the following prescription was adopted. 

i) We 
(re-)normalize all stellar spectra to the average flux in a 100~{\AA} passband 
around 5550 {\AA}, as in \citet{pic98}. This choice is not crucial for the different libraries at different resolution because we use the average flux within the band. 

ii) The spectra are 
then scaled with the luminosities (again using the average value in a 100~{\AA} 
passband centered at 5550 {\AA}) of theoretical spectra with the same stellar 
atmospheric parameters (interpolated as necessary) from the grid of 
\citet{lejetal97} -- which were the input of the M05 models. In short, the scaling of 
the empirical spectra is based on theoretical spectra, but our procedure allows us to keep, as much as possible, the continuum {\it shape} of the empirical spectra intact. For the MILES case we have also verified that the choice of width and wavelength is not crucial, by recalculating the scaling using a much larger $\lambda$~range, as adopted by other authors (see Section~5.3).

We note already now that the adopted procedure determines important features of the integrated stellar population model, a point on which we return when comparing our models with others from the literature (Section~5.3).

\subsection{Topping up missing cool dwarfs}
\label{sec:topdwarf}
Empirical libraries consistently lack cool dwarfs (precise temperatures naturally vary with metallicity, but rough guidelines are $<5000$~K for STELIB, $\sim4000-4500$~K for ELODIE, and $\sim4000$~K for MILES at non-solar metallicities), hence a supplement of theoretical spectra is necessary for completing the whole isochrone. 
For our purposes, the latest version of the MARCS library \citep{Gusetal08} was deemed the most suitable, comprising a comprehensive grid of cool stars (2500-8000 K) at very high resolution (R=20,000) over a large wavelength range (1300-200000~{\AA}). First, the standard chemical composition of these spectra follow the general trend of Milky Way stars as the empirical libraries, i.e. display a moderate enhancement of $\alpha$-elements ([$\alpha$/Fe]$=0.1-0.3$) at [Fe/H] between $-0.25$ and $-1.0$, with more metal-poor objects having [$\alpha$/Fe]$=+0.4$. 
\footnote{Attempting to keep the empirical approach intact, 
we also tried to take complementary stars from the Pickles library. We found, however, that the number of lower resolution Pickles stars needed to appropriately sample the lower MS, was 
enough to affect the overall resolution of the STELIB-based SEDs.}
Quite relevant, we also verified that the optical colours of MARCS stars and the corresponding integrated SSP models based solely on MARCS spectra are in good agreement with those based on empirical stars only (Section~\ref{sec:ssp}). This implies that our addition of these cool dwarfs does not alter the empirical spectral shape (around the $V$-band). We have also verified that the strength of integrated Lick indices is not affected by the use of cool MARCS stars, in spite some lines do not agree with the correspondent features as obtained from the extrapolations of MILES-based fitting functions (Johansson et al. 2010) into the very low-temperature regime. This happens as very cool dwarfs and giants ($T_{\rm eff}< 3500~^{o}{\rm K}$) have little effect on integrated optical line-strengths, unless the IMF is bottom-heavy (Maraston et al. 2003). It is well known that spectral lines in empirical and theoretical libraries do not always agree (e.g. Tripicco \& Bell 1995; Korn, Maraston \& Thomas 2005), but a proper comparison at very low temperatures has not yet been done. Since - as we mentioned - most empirical libraries lack very cool stars, one cannot fully trust that the extrapolation of the lines observed in warmer stars is correct. It goes beyond the scope of this paper to examine each spectral feature of the MARCS spectra in detail, and we defer this to a future publication.

Before entering the synthesis, the MARCS spectra were smoothed to the lower resolution of the empirical libraries. 

We have also calculated SSP models fully based on the MARCS library (Sec.~\ref{sec:nirexte}).

\subsection{Removing broken spectra}
\label{sec:ffcorr}

During the calculation of the MILES-based models a comparison to the value of $H_{\beta}$~from the fitting functions of \citet{jtm10} was performed, as we noted a discrepancy in the corresponding models. We found that indeed two stars, that otherwise do not display any peculiarity regarding the shape of their SEDs, had conspicuous $H_{\beta}$~strengths, one even had some Balmer emission. These were removed from the synthesis and replaced with stars of similar SED shape and normal Balmer strength values. The good agreement we find between solar-scaled models of Lick indices based on index-by-index fitting functions \citep{tmj11} and those calculated on the full SED presented in this paper comes from removing these anomalous stars (see Section~\ref{sec:lick}). This example supports our viewpoint that - when dealing with absorption lines using observations - it appears to be safer to take a look at the individual lines for spotting anomalous outliers, and the FF method is an effective way of doing so.

\subsection{$UV$-extension of the models}
\label{sec:uvexte}
For galaxy formation and evolution studies, an extension into the ultraviolet 
becomes necessary as the 
redshift gets progressively higher. Of the four sets of models, only 
the Pickles-based has a UV-extension. In order to improve on this situation, 
alternative versions of models have been constructed, where we have 
performed a concatenation with the SSP models of \citet{CMetal09a}, which 
are based on the fully theoretical UVBLUE/Kurucz library of high-resolution (R=10000) 
stellar spectra \citep{rodmeretal05}, and have a wavelength coverage of 
1000-4700~{\AA}. After downgrading the UVBLUE-based models to the appropriate 
resolution of each respective set of models, and normalising them onto 
the M05 models using the average flux in a 100 {\AA} wide passband centered at 
3050 {\AA}, the merging takes place around 3750 {\AA}. For the ELODIE models 
the merging is performed between 3960 and 3990~{\AA}.\footnote{The exact wavelength changes slightly for aesthetical purpouses.}

\begin{figure}
  \centering
    \includegraphics[width=0.49\textwidth]{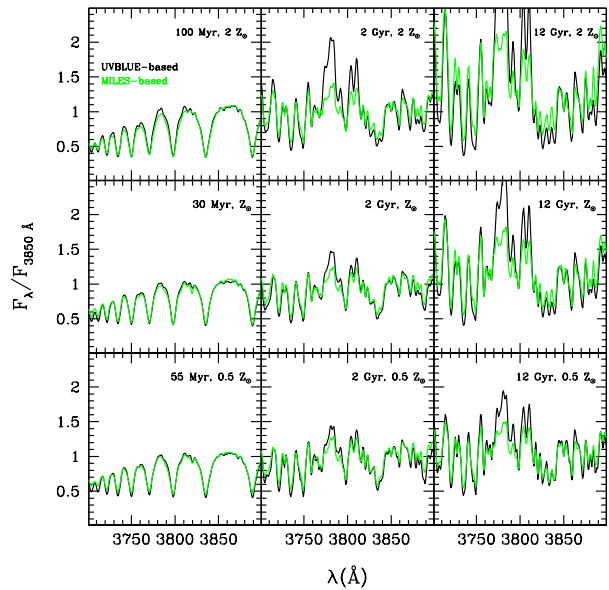}
    \caption{Zoom into the merging zone for MILES-based (green) and UVBLUE-based (black) models, for various ages and metallicities.}
    \label{fig:uvmerge}
\end{figure}
\begin{figure}
  \centering
    \includegraphics[width=0.49\textwidth]{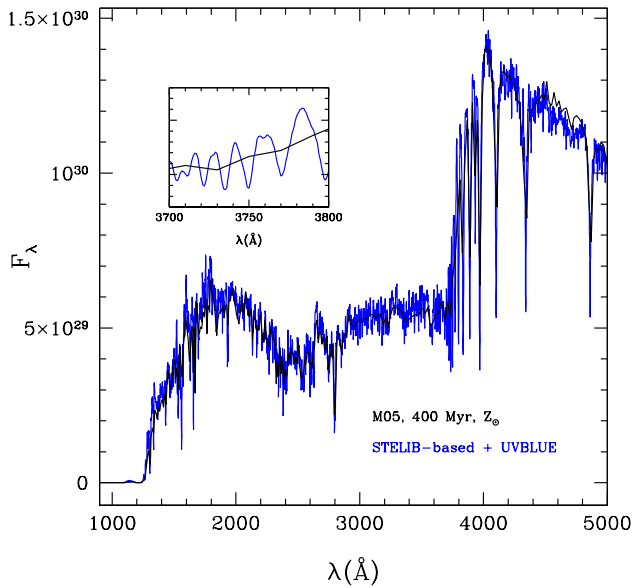}
    \caption{Comparison between the merged UBVLUE - STELIB 400 Myr SSP (blue) and the equivalent M05 model (black), and zoom into the merging region (inset).}
    \label{fig:uvmergem05}
\end{figure}

Figure~\ref{fig:uvmerge} zooms into the merging zone between UVBLUE-based and empirically-based MILES models for various ages and metallicities. The overlapping region looks excellent at young ages, where also the population flux in the UV is higher. At intermediate-ages the matching is still very good, though one notices some excess flux in the theoretical library especially at super-solar metallicity. At old ages the excess flux is evident at all metallicities. It is hard to say what the source of this noise is, but the flux level here is very low and should be irrelevant in spectral analysis. Finally, Figure~\ref{fig:uvmergem05} shows another example of merged models, using the 400 Myr STELIB-based model (blue) and the M05. The merging takes place here at 3752.5 \AA.

\subsection{High-resolution theoretical SSPs based on the MARCS library}
\label{sec:nirexte}
High-resolution models that reach further into the near-infrared than 
what can be accomplished with empirical spectra are equally useful. We therefore calculated a grid of SSP 
models adopting the MARCS library of high-resolution, theoretical stellar 
spectra \citep{Gusetal08}, which covers an impressive spectral range from 
1300~{\AA} to 20 $\mu$m with a resolution of R$=$20,000. The library is 
somewhat limited in terms of stellar temperatures -- the highest 
temperature is 8000 K at any metallicity -- which translates into a 
minimum age of 0.8 Gyr at 2 Z$_{\odot}$, and around 6 Gyr at the lowest 
metallicity. In the latter case a problem arises also for the oldest 
populations, where the shedding of mass towards the end of the RGB phase 
causes deeper, hotter layers in the stars to be exposed. If the mass-loss is 
severe enough, horizontal branch stars hotter than 8000~K will be produced. 
For a detailed account of the mass-loss prescription employed here, the reader 
is referred to M05. 

Furthermore, we calculated a version of the MILES-based SSP models in which we extended them into the UV as explained in Section 4.5 and into the near-IR using the MARCS-based models illustrated in this Section. 
These {\it merged} models are probably the models with the largest wavelength coverage at high resolution that are presently available.

\section{The SSP models}
\label{sec:ssp}
\subsection{Age and metallicity coverage as a function of the spectral library}
\begin{figure}
  \includegraphics[width=84mm]{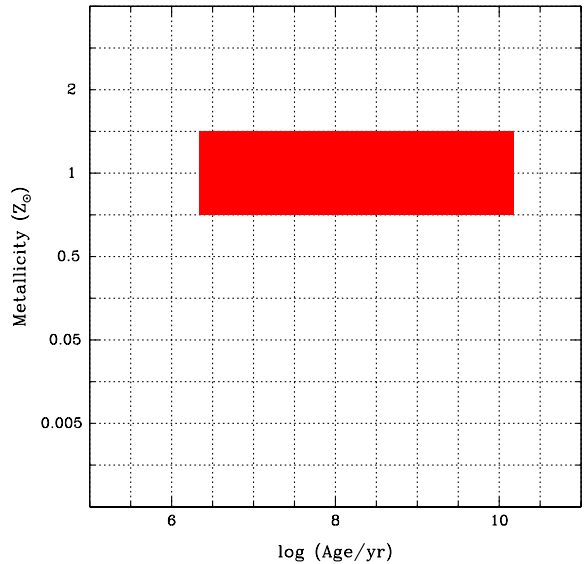}
  \caption{Schematic view of the available Pickles-based models, in 
    terms of age and metallicity, Z.}
  \label{fig:covpic}
\end{figure}

\begin{figure}
  \includegraphics[width=84mm]{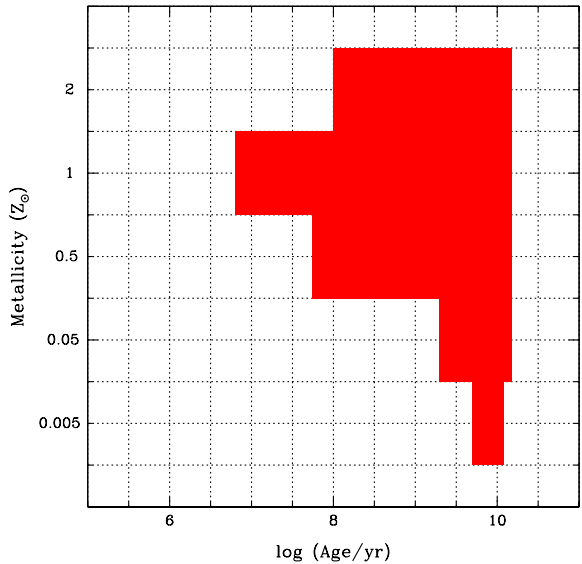}
  \caption{Same as Fig.\ref{fig:covpic} for the MILES library.}
  \label{fig:covmil}
\end{figure}

\begin{figure}
  \includegraphics[width=84mm]{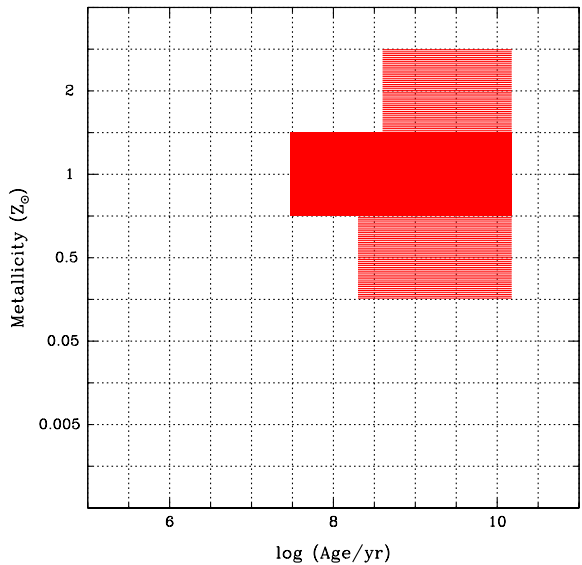}
  \caption{Same as Fig.\ref{fig:covpic} for the STELIB library. Models in the 
  striped area have limited wavelength coverage.}
  \label{fig:covste}
\end{figure}

\begin{figure}
  \includegraphics[width=84mm]{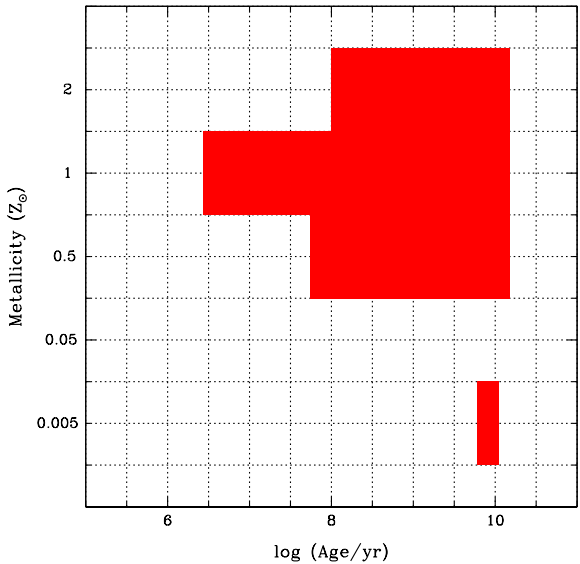}
  \caption{Same as Fig.\ref{fig:covpic} for the ELODIE library.}
  \label{fig:covelo}
\end{figure}

The simple stellar population (SSP) models that can be calculated for each of 
the four different empirical libraries, in terms of age and metallicity, are 
shown schematically in Figures~\ref{fig:covpic}-\ref{fig:covelo}. It is evident 
that empirical spectra alone cannot account for all desired combinations of 
these parameters. For example, no library contains the hot stars of low 
metallicity that are required to build metal-poor SSP models with young ages. 
Naturally, this is because star formation in the Milky Way takes place in the 
metal-rich disc.


As mentioned earlier, the Pickles-based SSP models (Figure~\ref{fig:covpic}) are 
only available at solar metallicity. The presence of O-type stars in the 
library allows for very young ages to be modelled (down to around 2 Myr, where 
the turn-off stars have a temperature of roughly 40000~K), but these spectra 
are also significantly attenuated by dust absorption in the UV. Hence the Pickles-based models 
with the youngest ages (below $\sim$100 Myr) should be used somewhat 
more cautiously. 

The MILES-based SSP models display the most comprehensive sampling of the 
model grid of all the four empirical libraries (see Fig.\ref{fig:covmil}). 
Reflecting the chemical evolution of the Milky Way, no particularly young ages 
can be modelled at the lowest metallicity. In general, the mapping of stellar 
atmospheric parameter at this metallicity is rather coarse, hence these models 
should be treated with some reservation. As previously mentioned, the 
[Fe/H]$=-$2.0 models, together with the [Fe/H]$=-$1.3 models, are expected to 
be enhanced in [Mg/Fe] by around $+$0.4 dex.

The amount of spectra in the STELIB library with twice solar, solar, and half 
solar metallicities is adequate to produce models down to relatively young 
ages. A caveat is that some of the stars of non-solar metallicity in the STELIB library do not have complete spectroscopic observations over the entire quoted wavelength range of $3200-9300$~\AA. The correspondent models will cover a narrower wavelength range of $3200-7900$~\AA\ (see Figure~\ref{fig:covste}).

The ELODIE library contains a number of very hot O-type stars, which in 
principle would allow us to produce models as young as 1 Myr. The spectra of 
these recently born stars are, regrettably but not surprisingly, severely 
reddened by close-in dust, and/or displaying various amounts of emission 
lines, and many of them had to be discarded in the population synthesis, 
thereby setting the age limit at solar metallicity to $\sim$3 Myr. At 
half-solar metallicity it is instead a lack of hotter supergiants ($>$ 6000 K) 
that sets a limit to the age of the youngest model. 

It should be noted, though, that despite the removal of the worst spectra, the 
youngest models are not completely unaffected by the dust reddening of their 
turnoff stars. At the lowest metallicity, 1/100 Z$_{\odot}$, the sampling of 
(T$_{eff}$, log(g))-space in the ELODIE library is fairly crude. A few ages 
can be modelled, but a complete evolutionary coverage is lacking, and they 
should thus be treated with some caution. There are not enough stars around 
\mbox{[Fe/H]$=-$1.30} to warrant the calculation of models of any age at that 
metallicity.

\subsection{The SSP spectra}
\subsubsection{MILES models at solar metallicity}
\begin{figure*}
  \includegraphics[width=\linewidth]{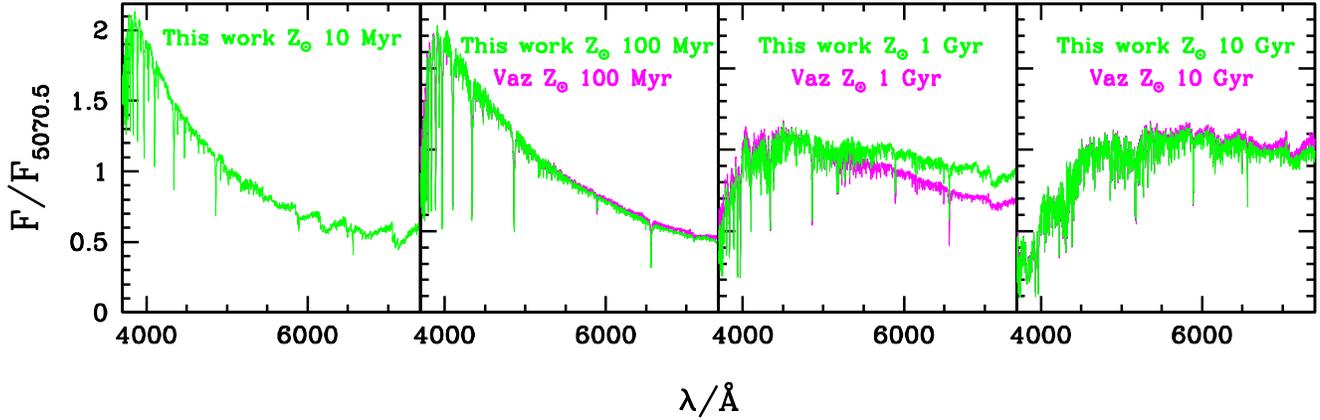}
  \caption{MILES-based models (green) for various ages and a solar metallicity, compared to models by Vazdekis et al. (2010, magenta). The latter models are not provided for 10 Myr of age. The higher flux longward of 5000 {\AA}~in our 1 Gyr model is due to the TP-AGB.}
  \label{fig:age_literature}
\end{figure*}
Figure~\ref{fig:age_literature} shows the age effects at fixed solar metallicity, comparing our MILES-based models to the models 
by \citet{vazetal10} (that are based on MILES). The 0.1 Gyr model compares well. The higher flux longward 5000 {\AA}~in the models at 1 Gyr is due to the TP-AGB phase. At old ages, the slightly higher flux longward 6000 {\AA}~in the Vazdekis models is due to the redder RGBs of the Padova tracks that are used in these models (see Figure 7 in M05 and discussion).

We also calculated broad-band colours, in order to spot differences due to the different libraries.
\begin{figure*}
  \includegraphics[width=0.7\textwidth]{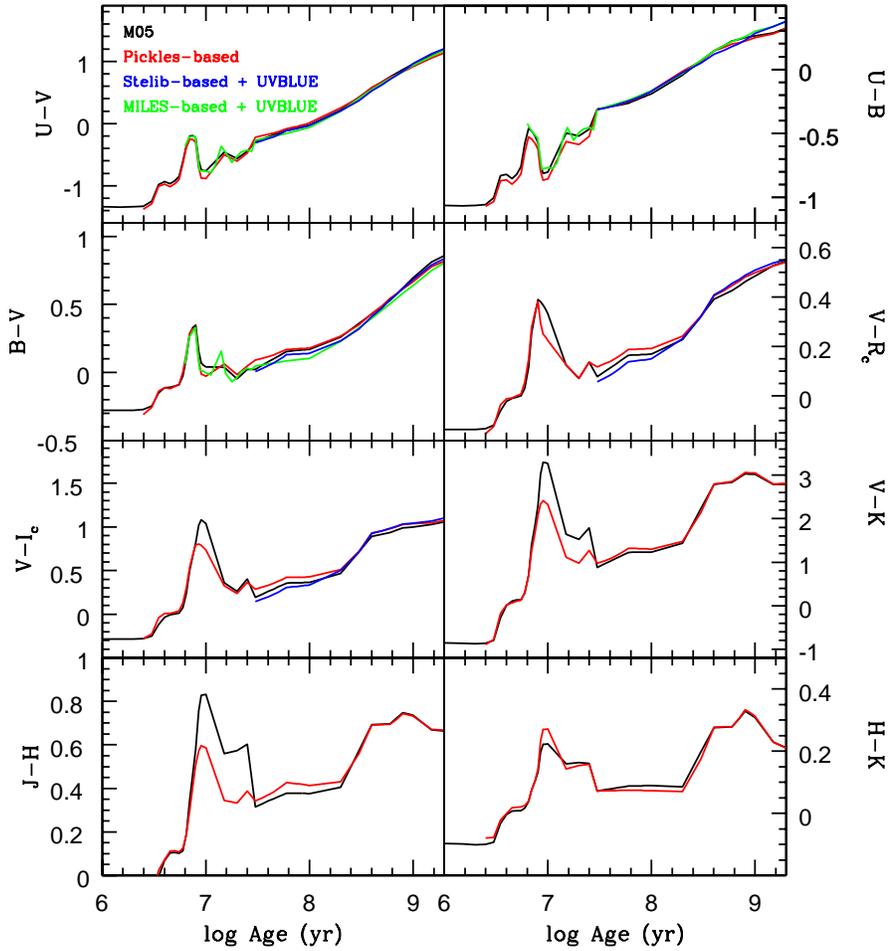}
  \caption{Integrated colours of young SSPs as a function of the various empirical spectral libraries. Colours plotted are those that could be computed given the wavelength extension of the library, see the text. The Pickles-based models using the M05-UVBLUE run identical and have been omitted.}
  \label{fig:colorsyoung}
\end{figure*}
\subsubsection{Young models as a function of spectral library}
Figure~\ref{fig:colorsyoung} displays the integrated colours of the youngest models as a function of the various spectral libraries. Colours plotted are those that could be computed given the intrinsic wavelength extension of the library through the near-IR, e.g. MILES allows the computation of the sole $B-V$, whereas STELIB allows also the computation of $V-R$ and $V-I$. $U-V$ and $U-B$ are calculated through the extension with the M05-UVBLUE models. The Pickles-based models based on their intrinsic UV from Pickles (which in turn is based on empirical spectra from the IUE database, cf. Section~3.1) and the version extended with the UVBLUE library run very similar, hence we show only the first ones. 

Overall, one sees an excellent agreement in the colours of very young stellar populations as a function of the spectral library for most colours. Some discrepancy is visible in the optical-to-near-IR colours of model ages where red supergiants exist ($t\sim~10~\rm Myr$), which are however well inside the typical scatter associated to colour indices at these young ages, which in turn is due to the short duration of the red supergiant phase (see Marigo et al. 2008 for a recent plot; see also Lan{\c c}on et al. 2008 for a general discussion on red supergiants)\nocite{maretal08,lanetal08}.

\begin{figure*}
  \centering
  \begin{minipage}{150mm}
  \includegraphics[width=140mm]{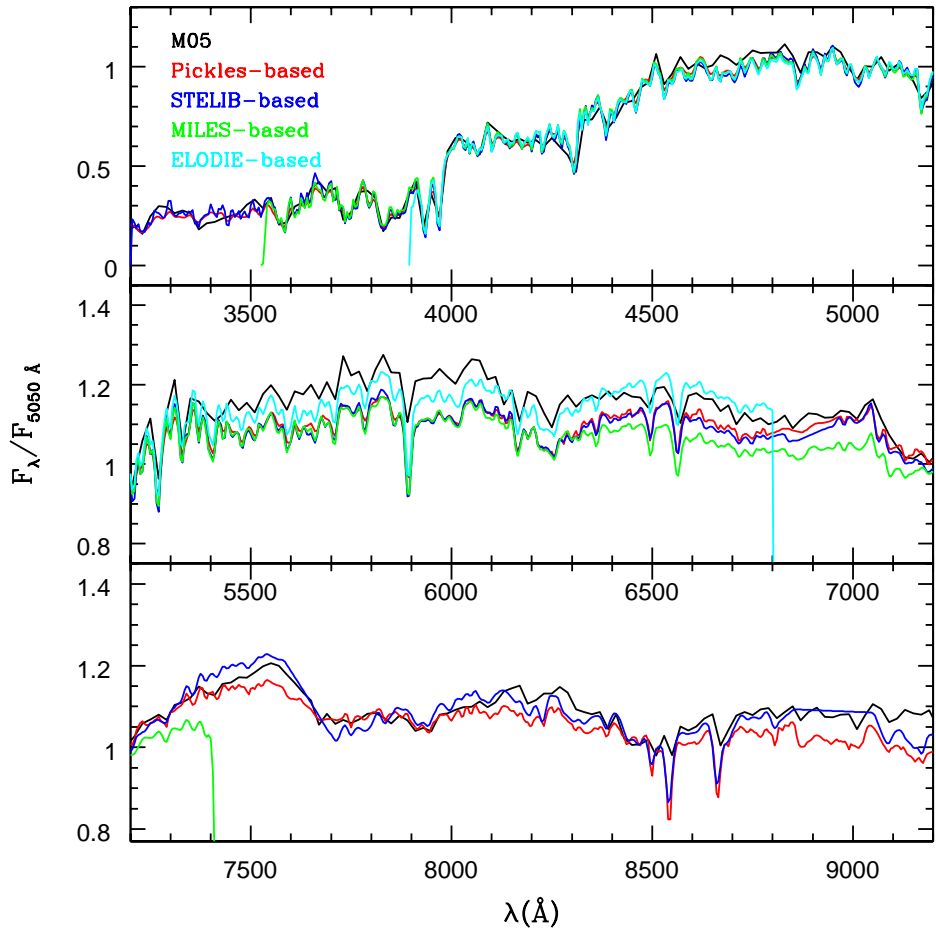}
  \caption{Comparison of the four sets of high-resolution empirical models and the original M05 
    models, normalised to 5050 \AA. SEDs refer to 12 Gyr, solar metallicity SSPs with RHB and a Salpeter IMF. Models have been smoothed to the quoted Pickles resolution.}
  \label{fig:12gyrsed}
  \end{minipage}
\end{figure*}
\begin{figure*}
  \includegraphics[width=0.7\linewidth]{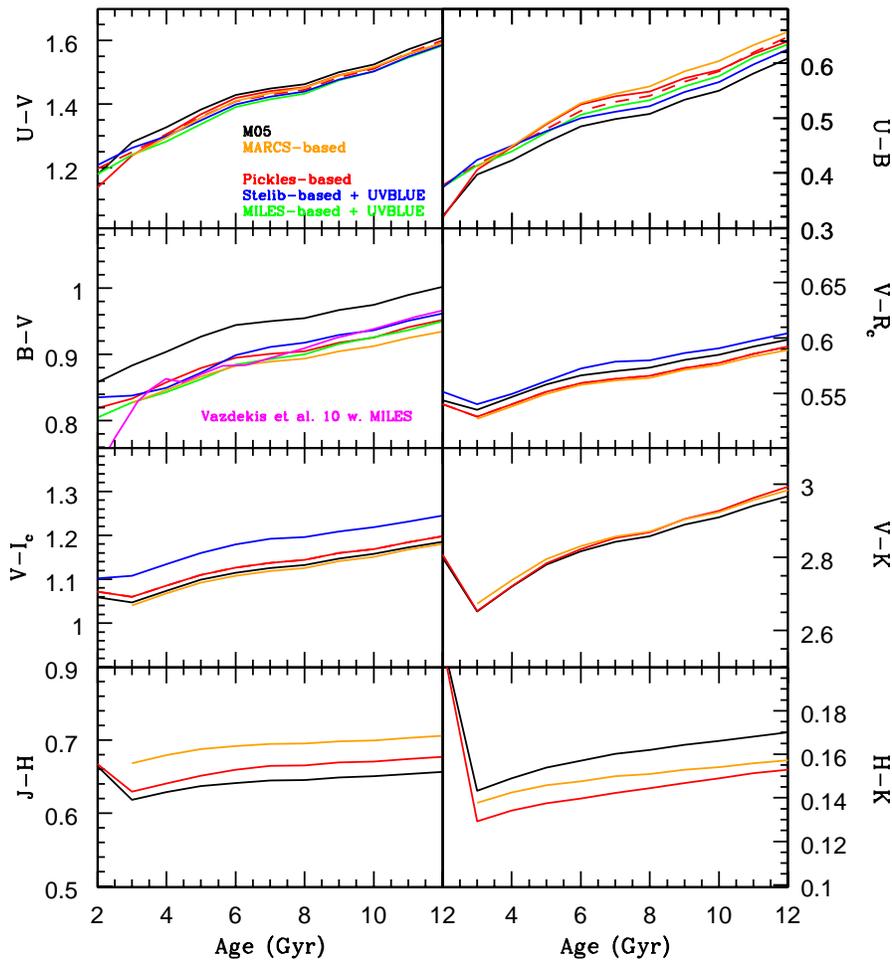}
  \caption{Integrated colours of old population models as a function of the various spectral libraries. The red dashed line refers to Pickles-based models in which their UV spectrum has been substituted with the theoretical UVBLUE spectrum (see Section 4.5). The models by Vazdekis et al. (2010) are also included.}
  \label{fig:colorsold}
\end{figure*}
\begin{figure}
  \centering
    \includegraphics[width=0.49\textwidth]{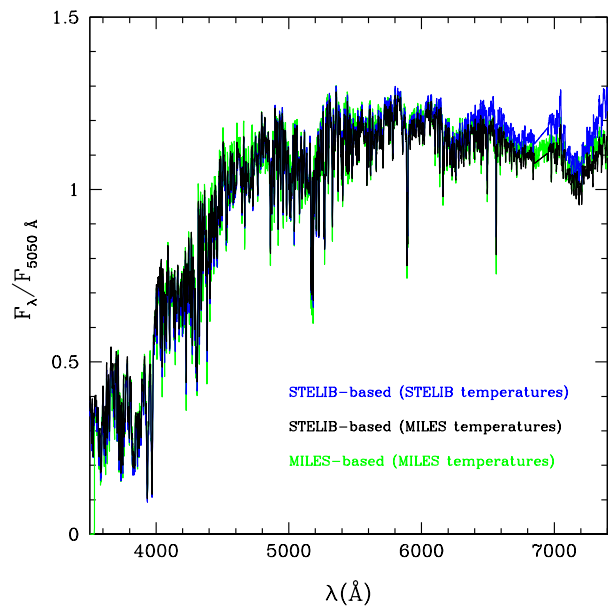}
    \caption{Effect of changing the temperature scale on stellar population models, for the illustrative example of a 12 Gyr solar metallicity SSP. STELIB stars have been tied onto the MILES temperature scale by using the fit of Figure~3. The original STELIB-based model (blue line) turns into one with lower flux in the near-IR (black line) when the MILES temperature scale is adopted. For comparison, the original MILES-based model is also plotted (green). The near-IR slope of the models depends substantially on the adopted temperature scale of RGB bump stars.}
\label{fig:tescalemilonstelib}
\end{figure}
\begin{figure*}
  \includegraphics[width=\linewidth]{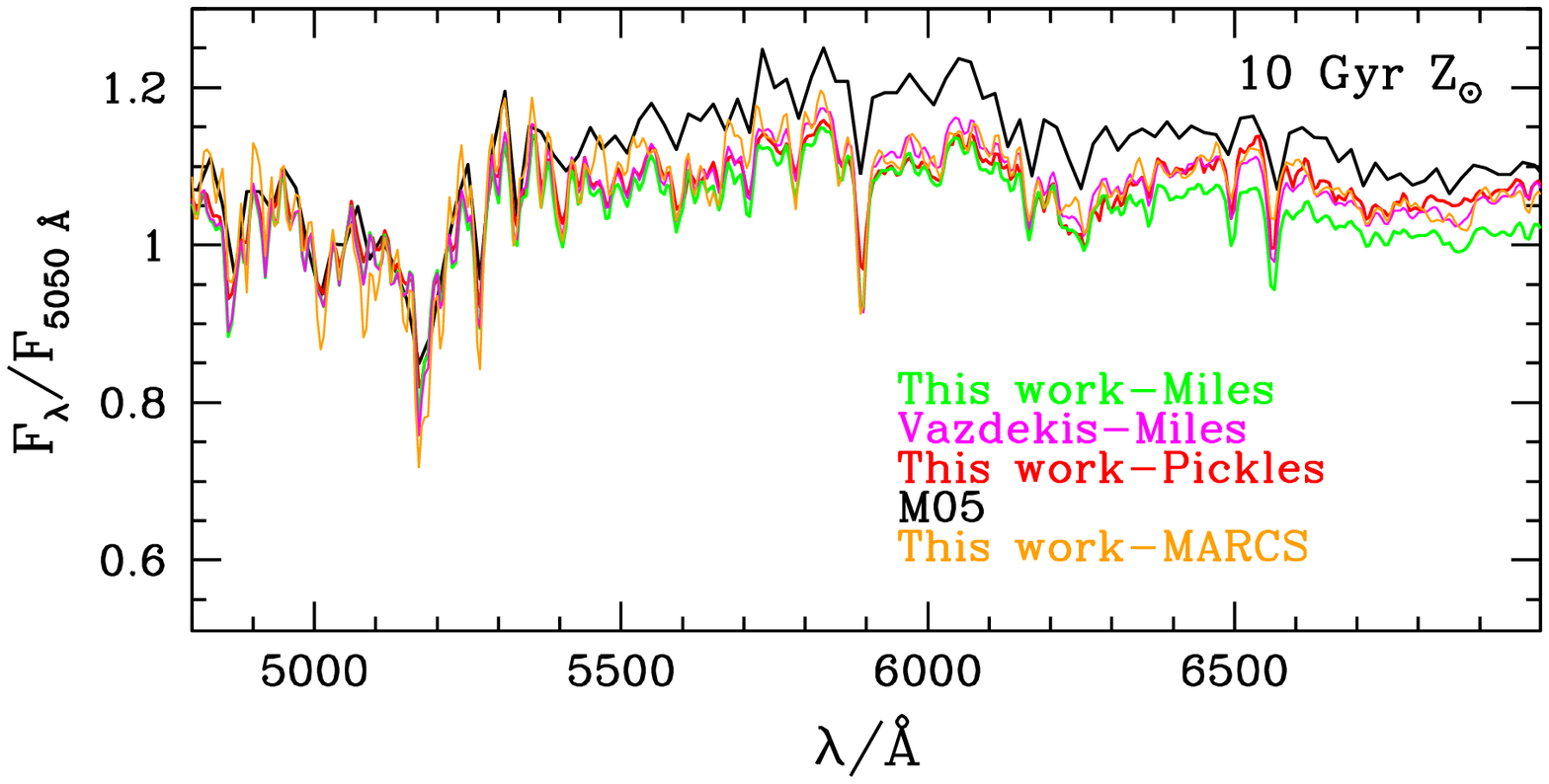}
  \caption{Remake of Figure~3 of \citet{CMetal09b}. The plot compares the spectral region around the $V$-band for different models, all referred to a 10 Gyr SSP with solar metallicity and a Salpeter IMF. New here is the inclusion of the models by Vazdekis et al. (2010) based on MILES (magenta line) and models based on the theoretical library of stellar spectra MARCS (orange line). All high-res. models have been smoothed to the Pickles's resolution.}
  \label{fig:lrg}
\end{figure*}

Note that the good matching for the continuum blue-ward of 4000~\AA\ supports our method of merging the empirically-based models with the theoretical UVBLUE-based models.

\subsubsection{Old models as a function of spectral library}
\label{sec:oldssp}
Figure~\ref{fig:12gyrsed} displays the overall spectral behaviour of the various models for an old, solar metallicity population. For clarity models have been smoothed to the quoted Pickles resolution. M05 is shown as a black line.

In general, significant differences are found between models based on different empirical spectral libraries.

Shortward $\lambda\sim~5000$~\AA\ (top panel) the models run very similar and close to the M05 model, the latter displaying a small excess flux around 4600~\AA. In the range $5200-6400$~\AA\ the various models display a very clear behaviour, i.e. all empirical models display a flux deficit with respect to the M05. We focus on this effect in the next subsection. The only exception to this trend are the ELODIE-based models, which have an overall higher flux, lying in between the other models and the M05, and reconciling onto the M05 at $\lambda>6300$. The reasons for such trend remain unclear, but it could possibly be related to the flux calibration, which at very high resolution is quite difficult.

In addition, as noticed in \citet{CMetal09a} and discussed in Section~\ref{libraries}, the MILES spectra lack flux longward of $\lambda \sim~6400$~\AA\, which is reflected in the SSP models (Figure~\ref{fig:12gyrsed}). 

We find that a substantial part of this effect is due to the assumed temperature scale for RGB-bump stars with temperatures between 4000 and 5500 $K$. In Section 3.6 we discuss the difference in temperature scales and we made an experiment in which we tied the STELIB library onto the MILES temperature scale. Here we show the resulting stellar population models in Figure~\ref{fig:tescalemilonstelib}. The blue line shows the original STELIB-based SSP, the black line the one we obtain using the MILES temperature scale, and the green line shows the MILES-based SSP. It is striking how much the near-IR flux has dropped by applying the simple re-scaling. 

We could figure out that this effect is dominated by RGB-bump stars.

This highlights the importance of understanding the effects of the assumptions on adopted temperature scales on stellar population models \citep{persal09}. For example, as discussed in \citet{CMetal09a}, the different slopes of the models affect optical-to-near-IR colours and the interpretation of galaxy data. 

We notice that our models based on the Pickles and STELIB libraries, but also those based on the theoretical library MARCS, in which no temperature scale calibration needed to be input, agree well long-ward 6000 \AA\ (cfr. Figure 15). Given the effect a different temperature scale for RGB stars has on the spectral shape of the integrated models, in Appendix \ref{sec:corrmil} we further expand on this point and provide a version of our MILES-based models in which we revise the near-IR slope to better match the one of the other libraries. Though we cannot be certain that the near-IR slope of MILES is not correct only because it does not compare well with those from other libraries, another fact in support of a revised slope is the observed-frame $r-i$~colours of massive luminous and red galaxies in the 2SLAQ survey, which were found to be systematically redder than colours obtained with a variety of existing models (Wake et al. 2006). In Maraston et al. (2009) we improved upon that issue using Pickles-based models, and we noticed that MILES-based models - with their intrinsic near-IR slope - did not perform as well (Figure 3,4 of Maraston et al. 2009). Though not being a clear-cut argument, it is an interesting hint towards a specific spectral shape. Obviously, observations of simpler systems such as star clusters would be very beneficial to constrain the effect.

Finally, at $\lambda\sim~6500$~\AA\ (middle and bottom panels) Pickles-based and STELIB-based models, and M05 run quite similar one to each other, except at 7500 \AA\, 8300 \AA\ and 9000 \AA\, where the Pickles-based models have consistently lower flux than the M05 models, whereas the STELIB-based models lie closer to M05. It is presently quite hard to calibrate these effects as one would need excellent quality near-IR spectra of old, solar metallicity GCs such as those in the Baade window, that are on the other hand strongly affected by reddening. While these will be available in the near future, we shall use galaxies to try discriminating among these effects in a future paper.

It should be stressed that in Figure~\ref{fig:12gyrsed} we have normalised all models at a certain wavelength (5050 \AA), which allows us to emphasise the spectral shape around it. This is obviously just a {\it visualisation}, the quantitative effect is derived by calculating the flux ratios of the models, i.e. by means of broadband colours.
These are shown in Figure~\ref{fig:colorsold}, which is analogous to Figure~\ref{fig:colorsyoung} for older ages. 

The colours of old models are contributed to by several stellar evolutionary phases (see e.g. M05) and the comparison of Figure~\ref{fig:colorsold} tells us that - at the same energetics - differences in the individual stellar spectra matter. This point was amply discussed in \citet{CMetal09a} where a first version of these models was published, and in M05 regarding MW GCs. 

In particular, we confirm the results of \citet{CMetal09a} for the $g'-r'$ colour, here for the $B-V$ colour, which when using empirical libraries - all of them - is bluer than the one relative to the M05 models, which are based on the BaSeL assembly of the Kurucz model atmospheres. In particular, notice that both the \citet{vazetal10} models and the M05-MARCS feature a bluer $B-V$, in close agreement with the empirical models.  This point will be amply discussed in the next subsection.

\subsection{The spectral shape around the $V$-band}

In \citet{CMetal09a} we found that SSP models based on empirical libraries (from Pickles, MILES, STELIB) displayed a lower flux around the $V$-band with respect to equivalent models based on the BaSeL theoretical library, an effect which helped to a better matching of the SDSS $g',r',i'$~colours of Luminous Red Galaxies around redshift 0.4. There, we also showed (Figure~4) that the same effect was found in the individual stars from the libraries, especially in giants stars, which secured us against possible  artificial effects introduced by the synthesis itself. The models used in that paper were a sub-sample of the extensive grid presented here, where we confirm our earlier finding\footnote{With the exception of the ELODIE-based models as noticed earlier}.

In this paper we consider two further models that strengthen our conclusions. Figure~\ref{fig:lrg} is equivalent to Figure 3 in \citet{CMetal09a}, and a blow-up of the middle panel of Figure~\ref{fig:12gyrsed} with two important addition, namely the \citet{vazetal10} models based on MILES (magenta) and our new models based on the theoretical stellar library MARCS (Section~4.6). The figure clearly shows that the completely independent \citet{vazetal10} models display exactly the same effect as ours. Furthermore, the M05-MARCS models behave as the empirically-based models. Note that the latter models do not require any adjustment or scaling of the MARCS spectra to be computed, which reinforces the conclusion that the procedure we adopt to insert the empirical libraries in the synthesis is appropriate. Finally, as we already pointed out, the slope long-ward 6000 \AA\ agrees very well between STELIB-based, Pickles-based, and MARCS-based models, while MILES-based models display lower flux, which is mostly due to the temperature calibration of RGB stars in this library.

\begin{figure*}
  \includegraphics[width=0.7\linewidth]{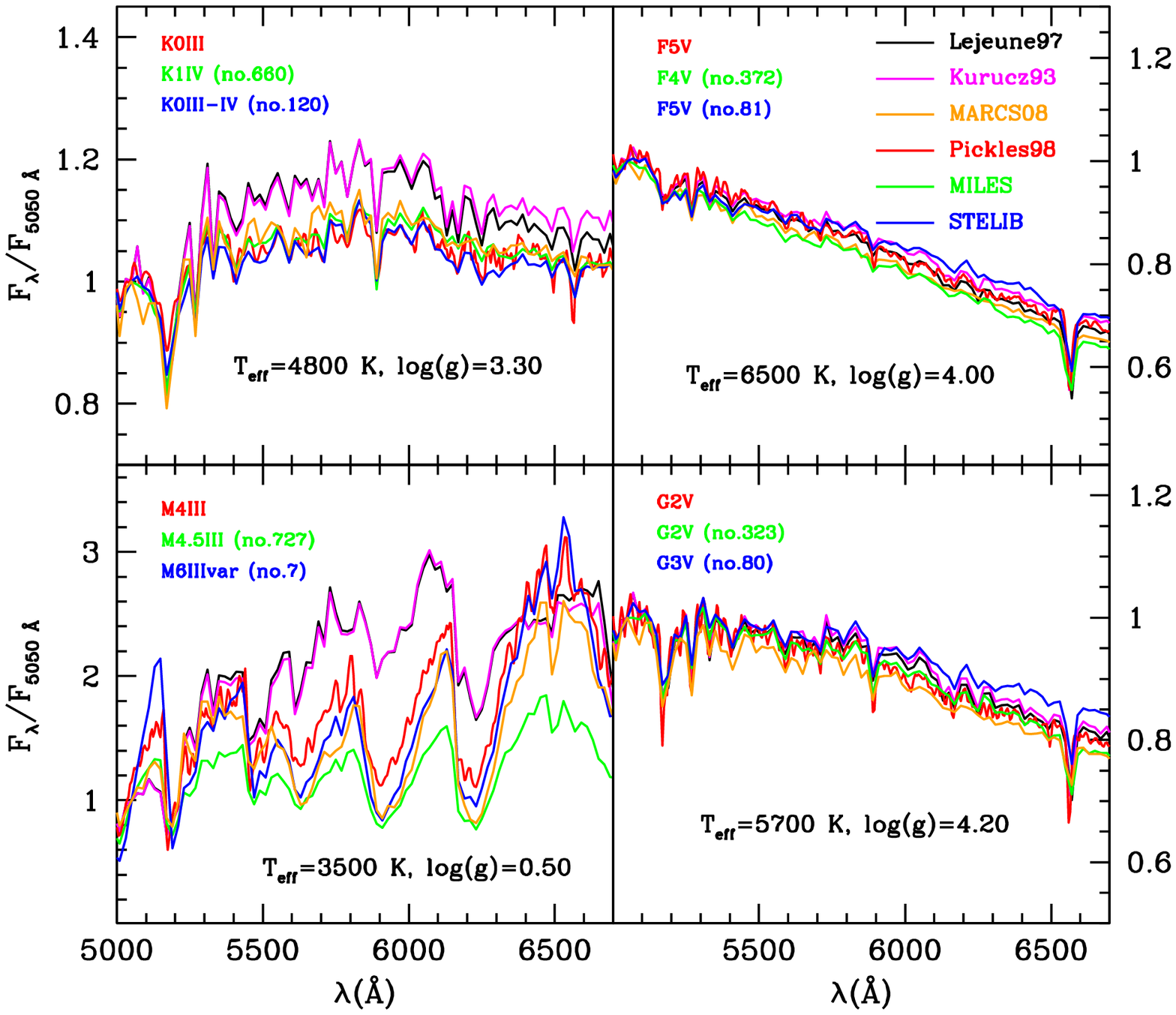}
  \caption{Individual stellar spectra from various libraries for several sets of stellar parameters (labelled in the panels). Remake of Figure~4 of \citet{CMetal09a} including now the MARCS library.}
  \label{fig:spectrastars}
\end{figure*}
\begin{figure}
  \includegraphics[width=0.89\linewidth]{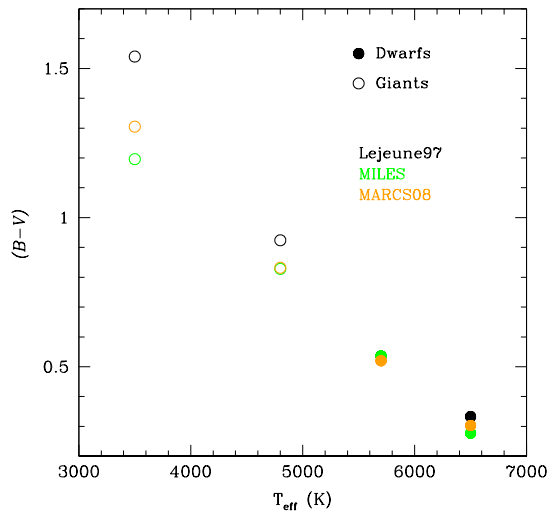}
  \caption{$B-V$~colours of individual stellar spectra from various libraries. The aim of this plot is to compare the optical colour between empirical libraries - using MILES as representative of them - and two theoretical libraries, namely the Lejeune et al. compilation of the Kurucz library and the MARCS library.}
  \label{fig:colstars}
\end{figure}

\begin{figure}
  \includegraphics[width=\linewidth]{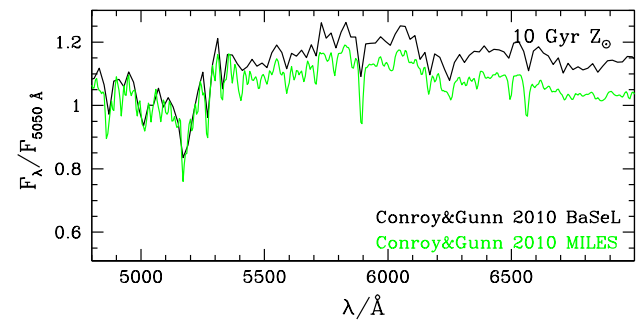}
  \caption{As Figure~\ref{fig:lrg} for the \citet{cg10} models (for a Chabrier IMF).}
  \label{fig:lrgcg}
\end{figure}
\begin{figure}
  \includegraphics[width=\linewidth]{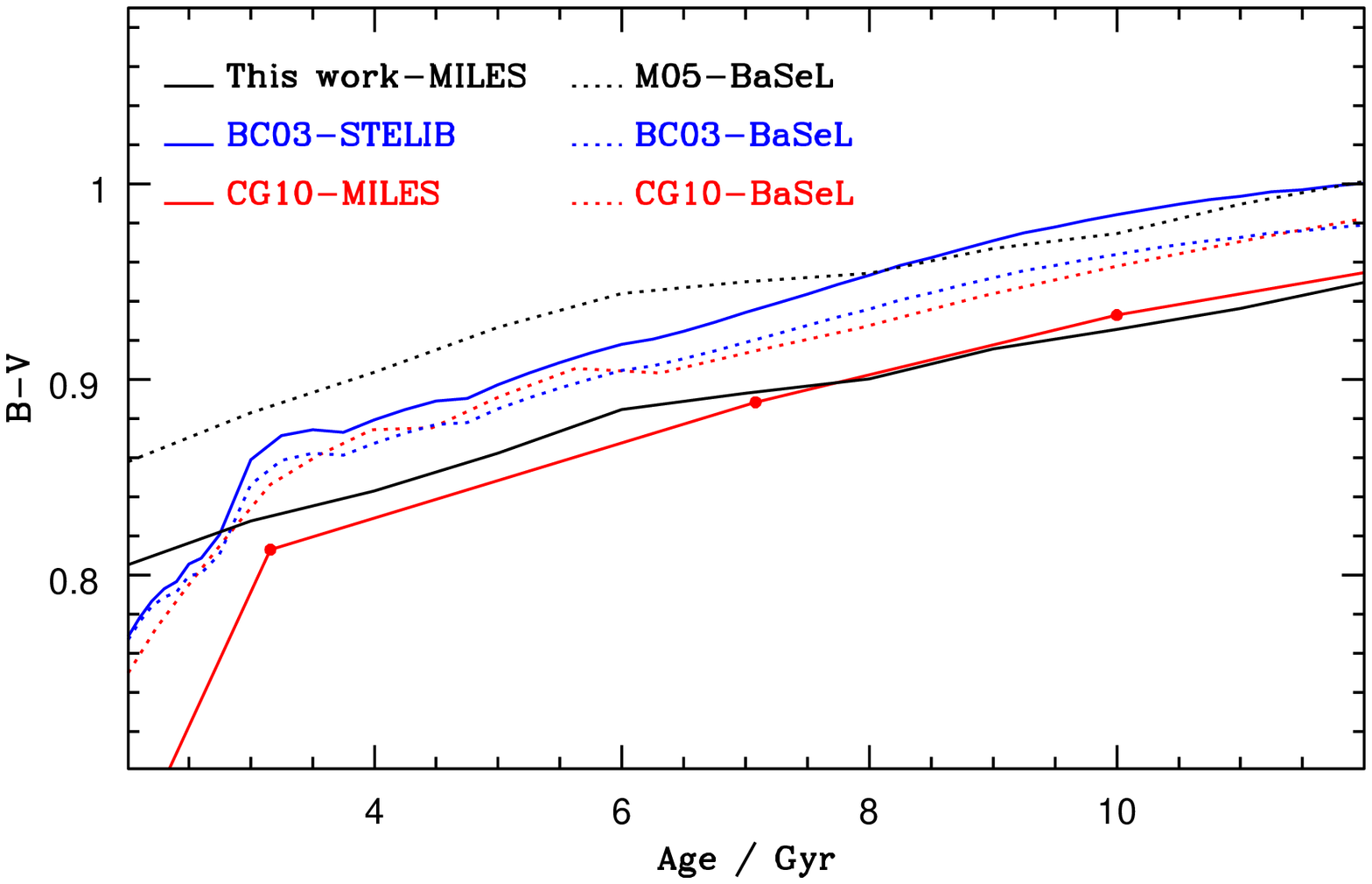}
  \caption{Integrated $B-V$ colour as a function of the spectral library, for different models, by Conroy \& Gunn (2010, red); Bruzual \& Charlot (2003, blue); M05 and this work (black). Solid lines refer to models based on empircal libraries, dotted to those based on the BaSeL library. The Conroy \& Gunn models refer to a Chabrier IMF, but this implies a minimum effect.}
  \label{fig:bvall}
\end{figure}

For completeness, we also show in Figure~\ref{fig:spectrastars} the individual stellar spectra from the various libraries, including the MARCS library, in a remake of Figure~4 of \citet{CMetal09a}. The MARCS synthetic stellar spectra behave as the empirical libraries at given stellar parameters. Finally, Figure~\ref{fig:colstars} shows that the $B-V$ colour of the individual stars from the empirical or the MARCS library is bluer than those of the Lejeune stars. 

Note that this result is obtained by trusting the stellar parameters provided by the libraries. The agreement with the MARCS-based modes suggests that this is the right procedure to adopt.

Figure~\ref{fig:lrgcg} shows the models of \citet{cg10}. These models behave as ours and the Vazdekis et al. models. This is at odds with the conclusions of Conroy and Gunn who claim to not recover the findings of \citet{CMetal09a}, i.e., they were unable to find any difference between the models based on the theoretical BaSeL library and the empirical MILES library, in particular, not on the $B-V$ colour. Figure~\ref{fig:bvall} shows the $B-V$~colour of their models, based on MILES and on BaSeL (red dotted and solid). Opposite to their statement, the colours are different, and in particular those based on the MILES library are bluer, as in our models (black lines in the same plot).

Conroy \& Gunn (2010) speculate that the effect described in \citet{CMetal09b} is due to the {\it "particular way in which the empirical stellar spectra were inserted in the synthesis by Maraston et al. (2009b)"} but as we clearly show in this paper, their models - in spectra and in colours - behave the same way as ours. 

Conroy \& Gunn (2010) also cite Bruzual \& Charlot (2003) for having reached their same conclusions on the $B-V$ colour. Instead, we show in Figure~\ref{fig:bvall} that the Bruzual \& Charlot (2003) models -  at odds with the cite by Conroy \& Gunn (2010) - also display an effect due to the stellar library, which has however the opposite sign, i.e. the STELIB-based colours are {\it redder} than those based on BaSeL. Here we note that these authors adopt a different procedure to insert the empirical spectra in the population synthesis, which is matching the empirical spectra to the theoretical ones (from BaSeL) in the $B-V$ colour, i.e. without using the library stellar parameters. This different procedure may be the origin of the effect, but is beyond the scope of this paper to investigate this issue in more depth.  

\subsection{The near-IR spectrum of old populations}
\label{nir}
 
\begin{figure}
  \includegraphics[width=84mm]{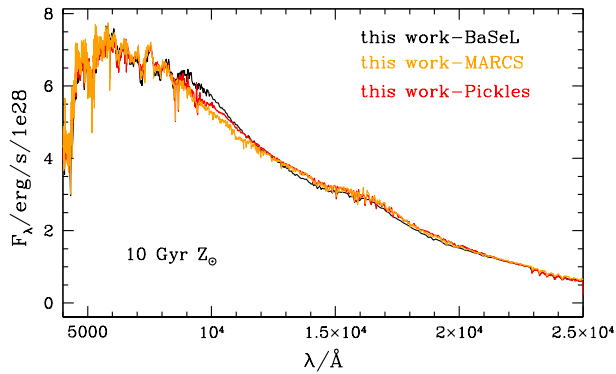}
  \caption{The near-IR spectrum of various SSP models, the one based on the empirical Pickles library (red), and two based on theoretical atmospheres, the standard M05 based on the BaSeL-Kurucz library (black) and the one presented here based on the MARCS theoretical spectra (orange). Models are 10 Gyr old and have solar metallicity.}
  \label{fig:nir}
\end{figure}

The Pickles-based models extend to the $K$-band, hence enabling us to check the near-IR wavelength regime of the spectra as a function of stellar library. We shall compare these with two variants of the M05 models based on theoretical libraries, the published M05 that was based on the BaSeL-Kurucz library, and the M05-MARCS that we publish here\footnote{Notice that both these models based on theoretical spectra  extend to much larger wavelengths, but here we are interested in the main emission from stars and in comparing with the Pickles-based models so we plot just up to the $K$-band}. Figure~\ref{fig:nir} shows the SEDs of 10 Gyr old populations according to the various models (we checked that the main features are invariant with the age). Though the overall shape is consistent, there are two regions where the models deviate from each other, namely around 1 micron, in correspondence to the $J$-band and close to 1.5 micron, in correspondence to the $H$-band. The behaviour is not simple, the models cross, with the standard M05 having more flux in the $J$-band and less around $H$, and viceversa for M05-MARCS and Pickles-based. These trends explain the difference in the corresponding broad-band colours. From the observational side the near-IR spectrum of galaxies and star clusters is becoming a new promising area of research 
\citep[e.g.,][]{rifetal07,rifetal08,lanetal08,siletal08,cesetal09,lyuetal10,rifetal11,minetal11} thanks to new instrumentation such as XShooter and others, and in preparation for the James Webb Space telescope. We shall soon be able to check these spectral details. 

\subsection{Lick indices}
\label{sec:lick}
A proper calculation and analysis of Lick absorption indices in integrated stellar populations require the account of element abundance ratios, because these lines are sensitive to their pattern, as amply discussed in the literature \citep[]
[,just to name a few among the most relevant works on the topic, see the latter paper for full referencing and discussion]{woretal92,davsadpel93,surben95,tribel95,gre97,traetal00,CMetal03,tmb03,tmb04,sch07,tmj11}. 

In particular, Figure 1 of \citet{CMetal03} clearly shows that the Mgb and Fe Lick lines of metal-rich galactic Bulge globular clusters with measured $[\alpha/Fe]\sim~0.3$~are not matched by stellar population models  
adopting the stellar index line-strengths of Milky Way solar-neighborhood stars, because the latter have a total solar metallicity and solar-scaled element ratios while the former have the same total metallicity, but $[\alpha/Fe]$~enhanced ratios. Star clusters in the Milky Way with abundance ratios known from individual stellar spectroscopy are not as ambiguous as distant galaxies and were used by Maraston et al. (2003) to definitively unveil the nature of the persistent discrepancy between Lick models and galaxy data. The models by Thomas et al. (2003), that account explicitly for various elemental ratios (see the same Figure), match well the Bulge globular cluster indices for $[\alpha/Fe]\sim~0.3$, consistent with the measurements from integrated-light spectroscopy. They also match the lines of massive galaxies.

The MILES library is not much different from the early Lick library from the point of view of abundance-ratios, in that it contains the same Milky Way-type of stars and in particular no Bulge stars, that would provide the metal-rich and $[\alpha/Fe]$-enhanced population that is missing in the solar vicinity. 

Hence, the indices calculated on the model SEDs we present in this paper are expected not to match the lines of metal-rich and $[\alpha/Fe]$-enhanced populations, that can be found in external massive galaxies and star clusters. For this reason, \citet{tmj11} calculated element-abundance-ratio-sensitive models based on MILES stars but following the same philosophy of Thomas et al. (2003), i.e. by theoretically manipulating the index response functions. 

\begin{figure*}
  \centering
  \begin{minipage}{180mm}
  \includegraphics[width=\textwidth]{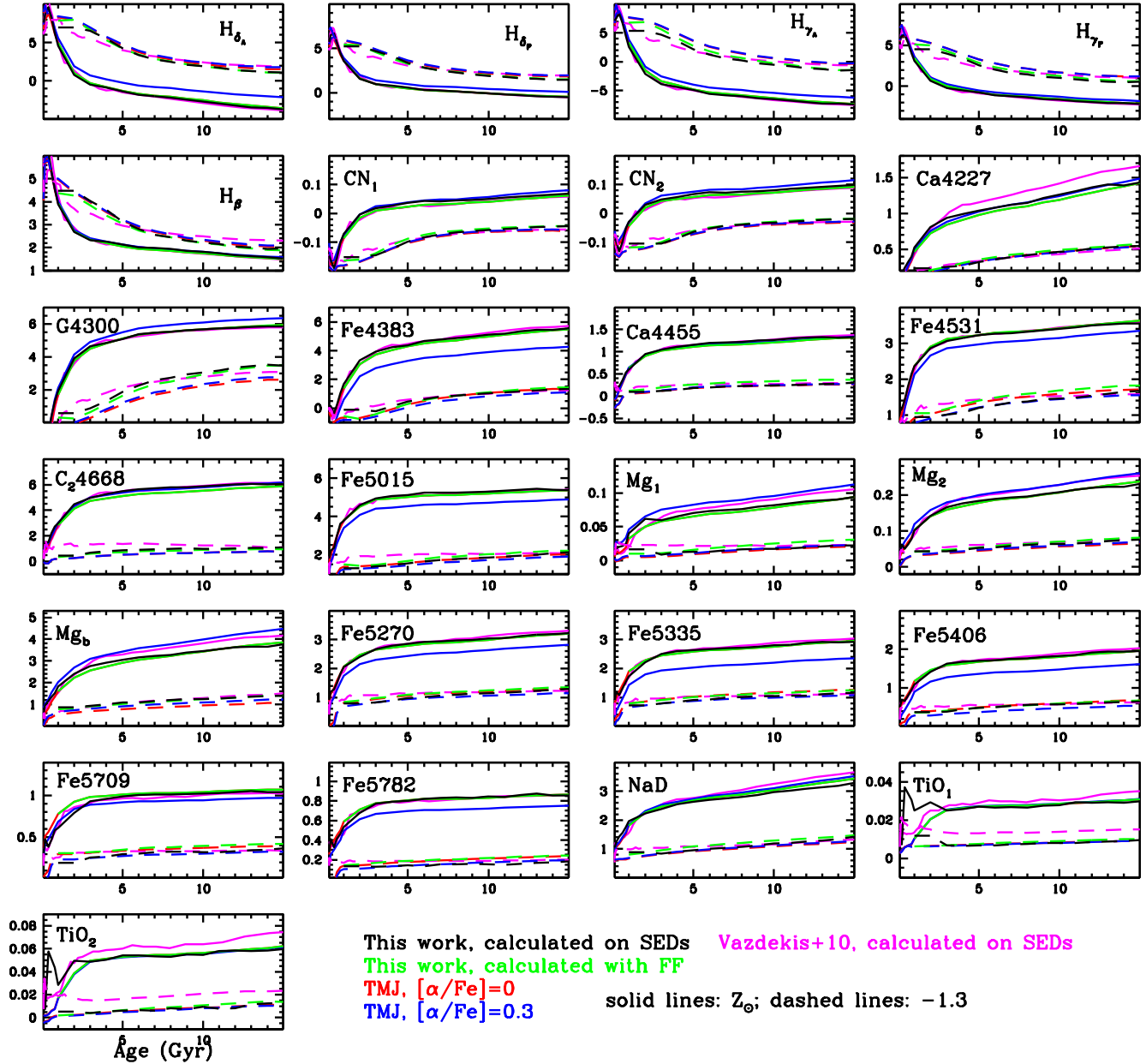}
   \caption{Comparison between different model Lick indices (at Lick resolution), namely: directly calculated on the MILES-based integrated SEDs presented in this paper (black lines); calculated through MILES-based FF (from Johansson et al. 2010, green line); from models with variable abundance ratios from TMJ for solar-scaled $[\alpha/Fe]=0.0$ (red) and $[\alpha/Fe]=0.3$~(blue). Solid and dashed line-styles refer to solar and sub-solar $\sim~-1.3$~metallicity. Vazdekis et al. (2010) models also directly calculated on SEDs and for the same metallicities are shown as magenta lines.}
  \label{fig:lickall}
  \end{minipage}
\end{figure*}

Nonetheless it is interesting to see how the Lick indices calculated on the integrated model SEDs of this paper compare with those from the TMJ models. Figure~\ref{fig:lickall} shows all 25 Lick indices in order of increasing wavelength from top to bottom, with the exception of the Balmer lines that are listed besides one each other at the top. Lick indices directly calculated on the model SEDs presented in this paper are shown as black lines, TMJ models for two different abundance-ratio patterns, namely solar-scaled (red) and $[\alpha/Fe]=0.3$~(blue), are also shown, with solid and dashed line-style referring to solar and sub-solar 
($\sim-1.35$)~total metallicity [Z/H].  Also shown as green lines are models obtained by integrating the MILES-based FF of Johansson et al. (2010). This type of models help us understanding whether discrepancies between the SED-based and the TMJ models are due to the SED vs FF approach. 

In general, indices from the integrated SEDs (black lines) should be in agreement with solar-scaled TMJ models at solar metallicity (solid red lines) and with $[\alpha/Fe]=0.3$-TMJ models at low metallicity (dotted blue lines), because of the intrinsic chemical bias Milky Way stars carry into the calculations (see discussion above). This is indeed the case for the vast majority of indices. Only a few cases do not follow this trend, on which we comment below. 

As mentioned, the comparison between green lines and black lines helps us spotting discrepancies due to the different approach, namely integration on SEDs or FF.
Overall, there is an amazing agreement between our calculations following the two approaches for most indices at both metallicities. Neglecting minor differences, such as e.g. the Ca4227, one index sticks out for having a discrepant value between direct integration (black) and FF approach (green), namely Fe4531 at sub-solar metallicity. We verified that this is due to a slight offset of the correspondent FF for giant stars.The Fe4531 indices of these stars are particularly sensitive to gravity effects (Johansson et al. 2010). Thus the modeling of this index is complicated, from both the SED and FF point of view, especially at metallicities below $-1.0$ where the number of MILES stars is fairly low. However, the chemical enrichment modelling by TMJ is able to rectify the offset introduced by the FF, such that the end-point TMJ $\alpha$-enhanced model (blue line) perfectly agrees with the values integrated onto the SEDs (black lines), and the TMJ index model is perfectly calibrated against GCs (see TMJ, Figure~3).

At sub-solar metallicity some indices do not follow the expected trend of matching the TMJ 
$[\alpha/Fe]$-enhanced (dotted blue lines). These are: $H\gamma$, which is weaker in the SED-based approach with respect to the TMJ model, and the G-band G4300 - and to a lesser extent Fe4383 and Fe5406 - which are stronger. In all these cases one sees that the use of FF in TMJ vs the SED-integration approach (i.e. green vs black) is not responsible for the effect (with the possible exception of $H\gamma$~at low ages, where the warmest MS dwarfs have a higher index value than the one approximated by the FFs). The major part of the effect comes from the correction of the element ratio bias in the TMJ approach. As discussed in TMB, these models assume that the stellar 'metallicities' [Fe/H] in the input library reflect true iron abundance rather than total metallicity. As a consequence, the metallicity scale gets slightly shifted through the bias correction, which leads to weaker metal and stronger Balmer indices at low metallicities. This effect is indeed seen here with G4300 and $H\gamma$ being most affected such that the index G4300 is weaker and $H\gamma$ stronger in the TMJ model. It should be noted that these particular indices follow a very good calibration with GC values as shown in TMJ.

At solar metallicity, there is none of such noticeable cases, and the SED-based calculations (solid black) and the TMJ models with solar-scaled abundance ratios (solid red lines) agree well. Some slight differences can be seen for Ca4227, $C_{2}4668$ and CN1, which ought to be ascribed to the FF being slightly lower than the SED calculation.

We note that the indices by Vazdekis et al. (2010) (magenta lines in Figure~\ref{fig:lickall}) - which are obtained through direct integration on their MILES-based model SEDs - display similar discrepancies for the above mentioned indices, but these are typically larger and show a different age dependence, see the G4300 band, $C_{2}4668$  or the Fe5015, among the others. Note also that some indices are well offset, such as the Ca4227 and the TiO indices. The exact reason for these difference is hard to assess, but it should be kept in mind that the Vazdekis et al. (2010) models are based on different evolutionary tracks (from Padova) which have redder RGBs than the tracks used in our models and a different overshooting (see M05).

Aside from the above discussion, note that the TiO indices are stronger at ages around 1 Gyr due to the TP-AGB contribution in the Maraston models and the use of the empirical TP-AGB spectra (see Section~3.7)\footnote{The MILES library does not contain TP-AGB star spectra hence these are not included in the TMJ fitting-function-based models}. Also the Mg$_{\it 1}$ index, which is known to be sensitive to Carbon (Thomas, Johansson, Maraston 2011; see also Tripicco \& Bell 1995), is stronger at the same ages in the SED-based models. 

\section{Testing the models with globular clusters}
\label{sec:testing}
In this section we provide a few examples on how the new models perform in fitting real objects. Following our previous works, we use globular clusters as they are the closest approximation to simple stellar populations, i.e. age and metallicity for each cluster can be determined from resolved studies independently of stellar population models. Suitable for this test are Milky Way and Magellanic Clouds objects. As we are mostly interested to test the effect of spectral libraries, which - as they do not contain any TP-AGB star  - is mostly concentrated on old populations, we shall not repeat all comparison plots published in M05 using Magellanic Clouds clusters. In the following we shall use only MW GCs data. As is well known, MW field and globular cluster stars feature a well defined pattern of chemical abundance ratios, in particular the $[\alpha/Fe]$~ratio -  at given total metallicity.  This introduces a {\it bias} in models based on empirical MW stars (Maraston et al. 2003; Thomas, Maraston \& Bender~2003) that we shall take qualitatively into account in our comparisons. Note that empirical libraries contain - besides field stars - also globular cluster stars, in particular bright red giant branch stars, but these show a similar element abundance pattern as field stars \citep{prietal05} hence their inclusion does not affect our conclusions.

In the following we show tests involving both GC spectra as well as GCs broad-band colours.
\subsection{Globular cluster spectra}
We use the samples of \citet{puetal02} and \citet{ricardoetal05}, with main focus on the latter, which 
consists of 40 objects that represent the full range of MW GC 
parameters at an instrumental resolution of 3.1 {\AA} FWHM, and S/N ratios 
between 100 and 400. \citet{puetal02}'s spectra have a coarser
resolution (6.7 {\AA} FWHM) because the aim was to evaluate Lick indices, and have concentrated on the most metal-rich Bulge clusters, but 11 out of their total number of 12 objects are in common 
with the \citet{ricardoetal05} compilation, thus enabling a comparison also between 
different samples (see Section~6.1.3). Both sets of spectra have been flux calibrated, but neither 
have been corrected for reddening.

GC ages and metallicities from our new SED models have been derived through a standard
SED-fitting procedure, where we compute the $\chi^2$ between observed and model SEDs over the wavelength range 3600-6200 {\AA}, pixel by pixel (2600 pixels), using the S/N spectra for assigning errors to each pixel.\footnote{The Schiavon  
et al. spectra are corrupted around 4546~\AA~and 5015~\AA~and this range has been masked out.} The observed spectra are de-reddened prior to the fitting, using a simplified 
version of the wavelength-dependent MW interstellar extinction law of 
\citet{fit99} with R=3.1 and the $E(B-V)$~parameter compiled in \citet{ricardoetal05} 
This parameter depends strongly on the line-of-sight; however, since the purpose of this exercise is to show 
that, despite a rather simple approach, very reasonable estimates of GC 
parameters can be obtained in all but a few cases, such approximations 
may be forgiven.

For testing we use both the MILES-based and STELIB-based models, which have a resolution comparable to the GC spectra.
Only the MILES-based models could be used for all GCs, as they exhibit a wide enough metallicity coverage (cf. Figure~7). Model grids have been interpolated in age and metallicity in order to avoid clustering of solutions around the grid points. The Horizontal Branch (HB) 
morphology has been taken into account, in the sense that model SEDs have been pre-selected according to 
it\footnote{The models of this paper as the M05 models are provided for two morphologies of the Horizontal Branch, namely intermediate-red and blue, following a standard recipe for mass-loss that was calibrated on Milky Way GCs. See M05 for details.}. SED fits have been performed using both Salpeter and Kroupa (2001) IMF-models. Results will be compared 
to CMD-derived ages, by \citet{deangetal05} and \citet{marfraetal09} and metallicities from the latest 
version of the \citet{har96} catalogue of Milky Way Globular Clusters and from Marin-Franch et al. (2009).  

\begin{figure*}
  \centering
  \begin{minipage}{150mm}
    \includegraphics[width=140mm]{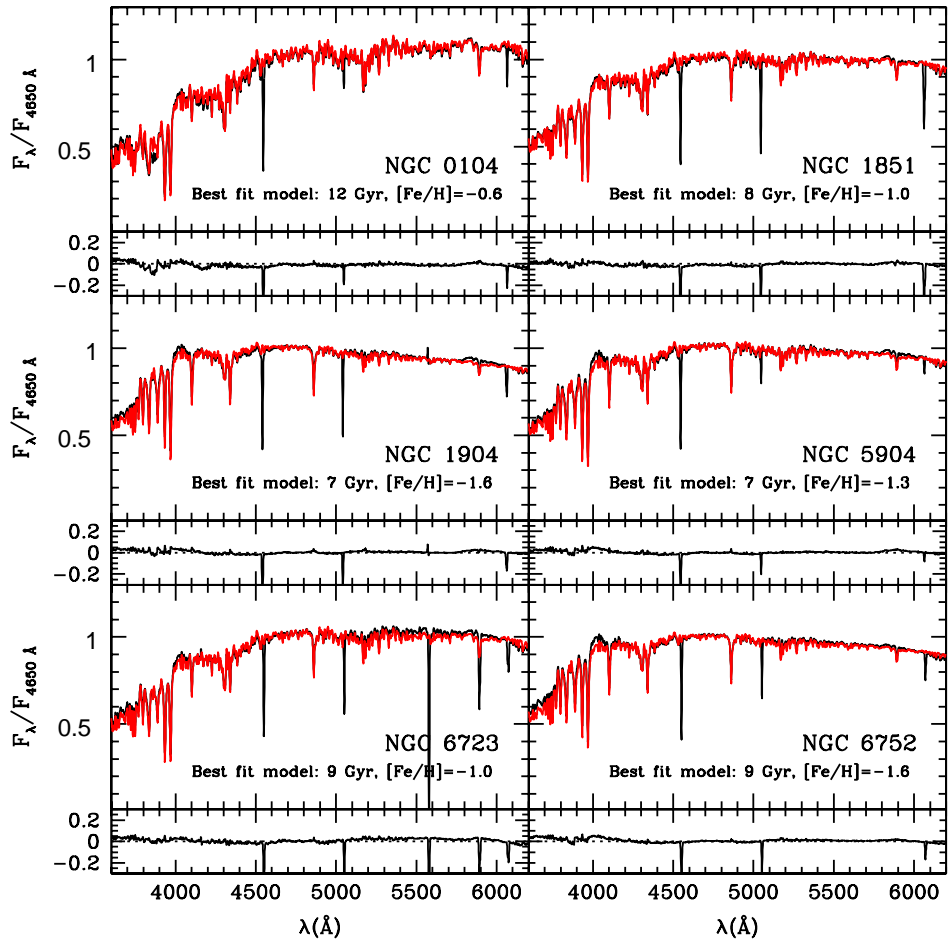}
    \caption{GC SEDs from \citet[black]{ricardoetal05}, overplotted 
      with their best-fitting, in terms of minimum $\chi^{2}$, MILES-based 
      template (red), for a Kroupa IMF, taking into account the HB 
      morphology of each cluster. The age and metallicity parameters of the best fits are indicated and the residuals are given below 
      each panel.}
    \label{fig:gcseds}
  \end{minipage}
\end{figure*}

\begin{figure}
  \includegraphics[width=90mm]{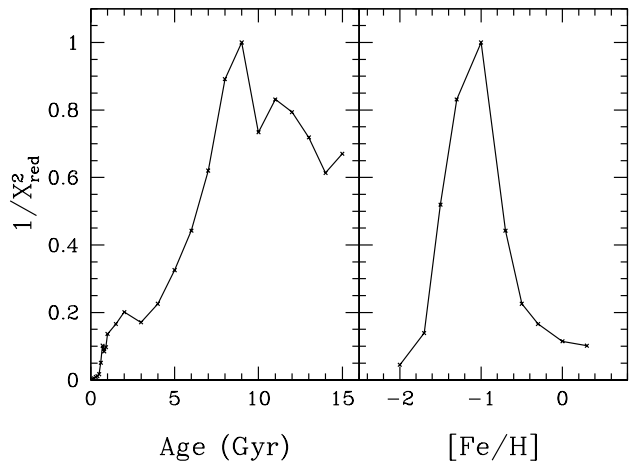}
  \caption{Marginalized age- and metallicity $\chi^2$ distributions for the 
  \citet{ricardoetal05} globular cluster NGC~1851 from fitting with MILES-based 
  models, normalized at the peak values. This is one of the 
  better fits.}
  \label{fig:gcchigood}
\end{figure}

\begin{figure}
  \includegraphics[width=84mm]{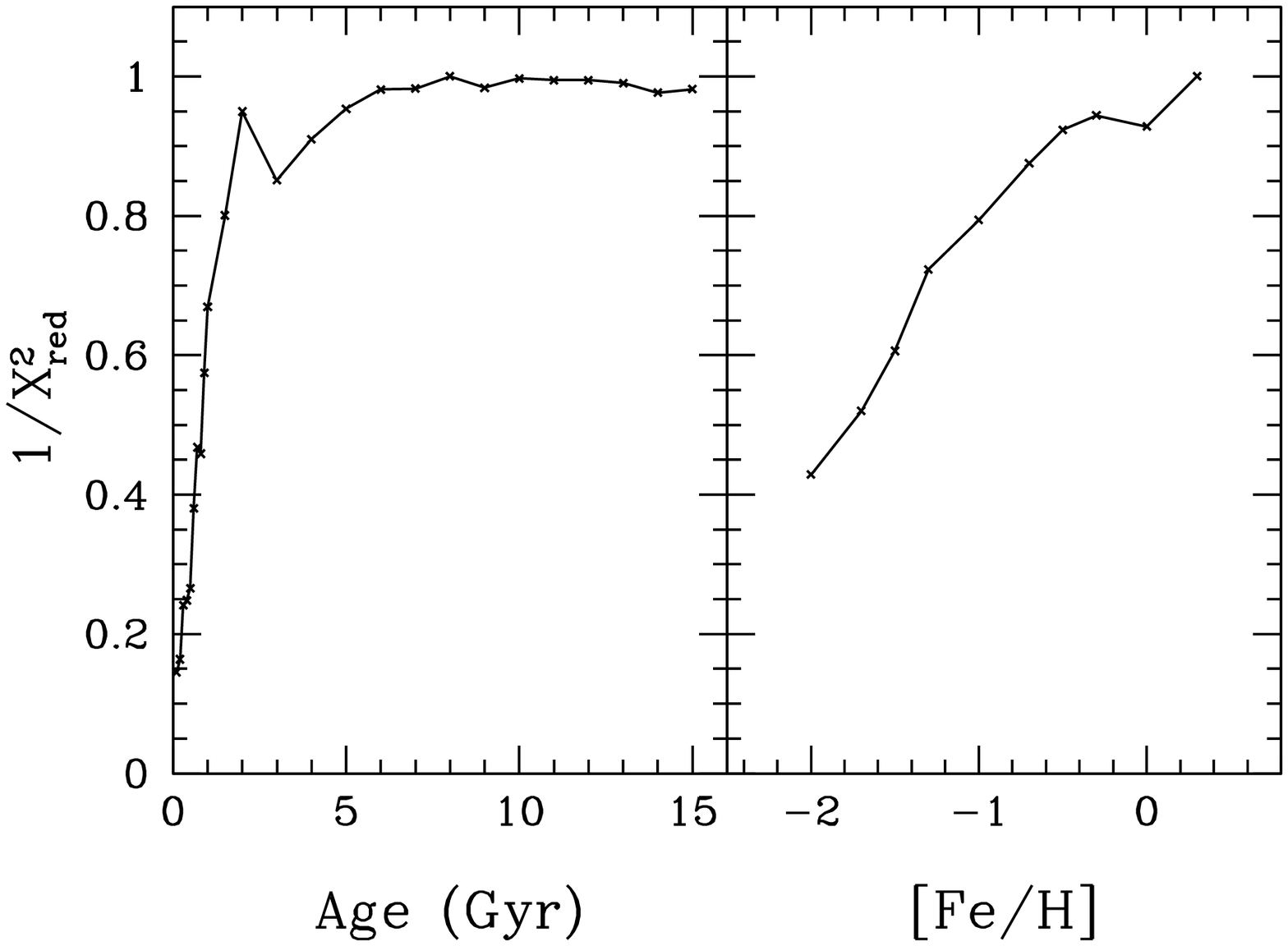}
  \caption{Same as Fig.\ref{fig:gcchigood}, but for NGC~6544. Notice the large 
    age-metallicity degeneracies, making this clusters one of the most poorly 
    constrained.}
  \label{fig:gcchibad}
\end{figure}

Figure~\ref{fig:gcseds} shows examples of SED fits of GCs, with the best-fitting MILES-based models shown in red. The six clusters represent a fair sampling of 
the MW GC parameter space, and are also those with the least amount 
of reddening (E(B-V)$\le0.06$). In general, the fits are excellent and the parameters of the best fits are 
in good agreement with the age values derived in \citet{deangetal05} and the metallicities from 
\citet{har96}\footnote{In considering these comparisons it is important to remember that we show the peak 
values in the marginalised $\chi^2$~distributions. Using a 68 \% confidence interval 
we obtain very large age and metallicity ranges -- a perfect showcase for 
the age-metallicity degeneracy.  On the other hand, in most cases the peak values are in excellent agreement with the CMD determinations. Two such examples are given in 
Figures~\ref{fig:gcchigood} and~\ref{fig:gcchibad}, where the normalised age- 
and metallicity probability distributions are highlighted for one of the best 
(NGC~1851) and one of the worst (NGC~6544) fits. For NGC 1851 we obtain 9.0 Gyr and [Fe/H]=-1.0 vs
CMD values of 9.0 Gyr and [Fe/H]=-1.2, with a confidence interval of  0.7-15 Gyr, and -1.7 to +0.3.
For the highly-reddened cluster NGC 6544 literature values are 9.4 Gyr and [Fe/H]=-1.56, and we get 8.0 Gyr and [Fe/H]=+0.3 with a 68\% confidence interval of 1.5-15 Gyr, and -1.0 to +0.3.}. 

As pointed out by Vazdekis et al. (2010), the discrepancy around 
3850~{\AA} visible for NGC~104 (47 Tuc) is due to the 
strong CN features of this cluster compared to the stars composing the libraries. In \citet{tjm11} it is shown that the corresponding CN Lick indices of such GCs can be matched by enhancing Nitrogen.

We also spot a recurring discrepancy between 4000 and 4100~{\AA}, where the model flux is slightly weaker compared to the observed spectra.
Finally, the models display somewhat lower flux levels blueward of $\sim3750$~{\AA}, 
but here the S/N has decreased roughly a factor~3 compared to the values 
around 5000~{\AA}.
A similar comparison is provided by Vazdekis et al. (2010) for the clusters NGC~104, 5904 and 7089 (their Figure~20), with derived cluster parameters in good agreement with ours (also for NGC~7089, not shown here).

So far we have focussed on the comparison of the overall SED. We do not attempt a detailed comparison of individual absorption features, because the latter requires the account of element abundance ratios in the models, which is done in a series of companion papers\footnote{Spectral features at other wavelengths for the present models will be addressed in future works.}.
Thomas, Maraston \& Johansson (2011, hereafter TMJ11) present the flux-calibrated version of the widely-used Thomas, Maraston \& Bender~(2003) model Lick indices with variable abundance ratios. These have been upgraded using empirical calibrations of the index strength as a function of stellar parameters \citep[fitting functions,][]{jtm10} based on the MILES library. \citet{tjm11} address in detail the comparison of these latest model Lick indices with Milky Way globular cluster data. 

\begin{figure}
  \includegraphics[width=0.49\textwidth]{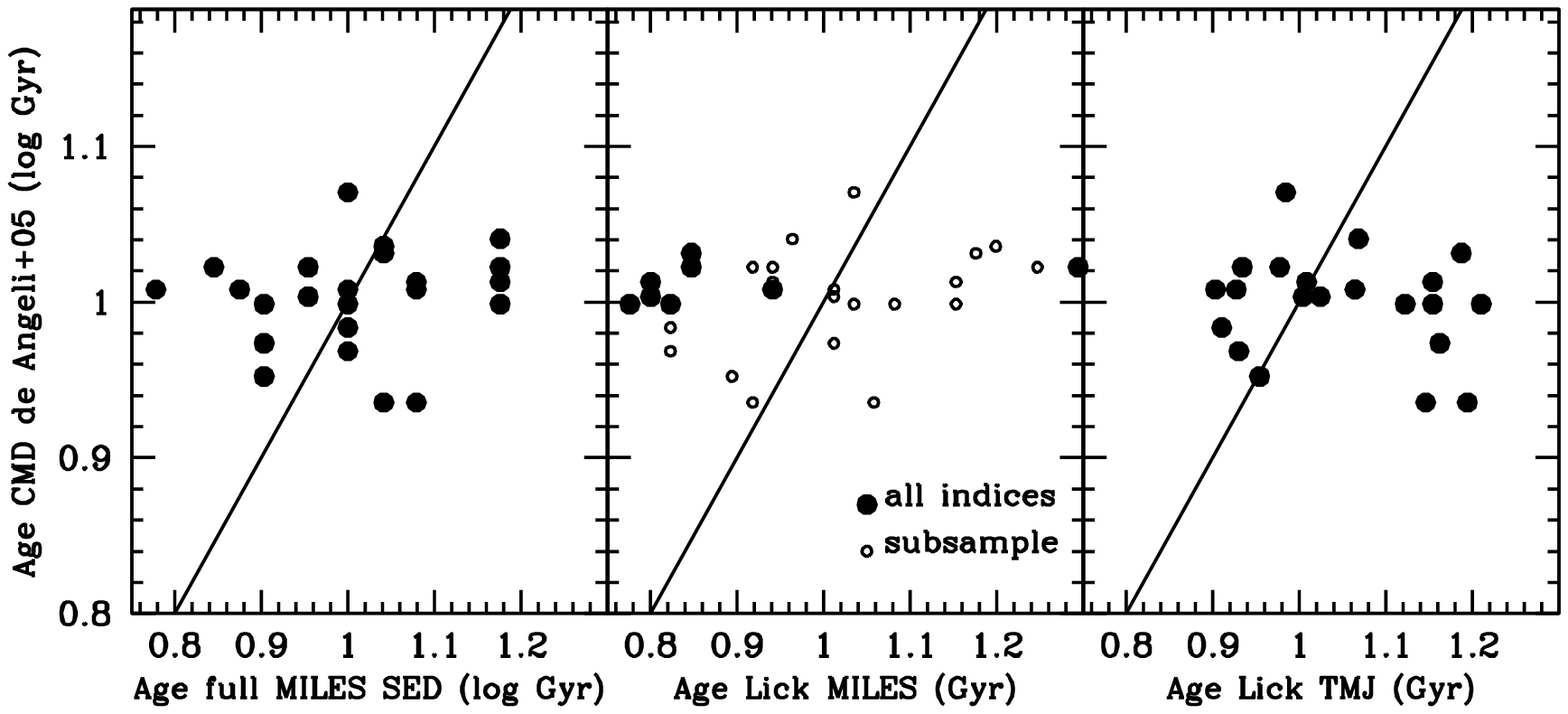}
  \caption{Comparison of ages obtained for the Schiavon (2005) GC sample as obtained via (panels from left to right): a) full SED fit on MILES-based SEDs (see Figure~\ref{fig:gcseds}); b) Lick absorption line indices calculated on the same MILES-based SEDs of case a), using: all available indices (filled symbols); a selection of indices the combination of which is little sensitive to element abundance ratios (TJM11) ; c) the TMJ Lick index models based on MILES with variable abundance ratios. On the $y$-scale we plot ages derived from CMD fit from De Angeli et al. (2005).}
  \label{fig:mastrolickagedeangeli}
\end{figure}

\begin{figure}
  \includegraphics[width=0.49\textwidth]{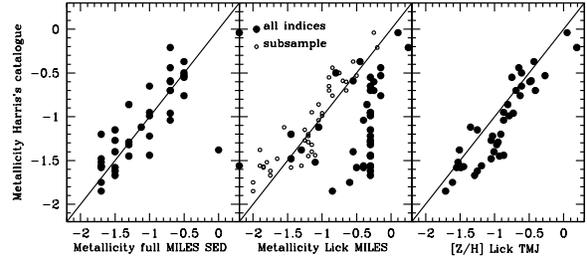}
  \caption{As in Figure~\ref{fig:mastrolickagedeangeli} for the metallicity. On the $y$-scale we plot values from the Harris (1996 and updated versions) catalogue, which are held to reflect the [Fe/H] parameter. For the TMJ models we plot the total metallicity [Z/H].}
  \label{fig:mastrolickmetalsdeangeli}
\end{figure}

\begin{figure}
  \includegraphics[width=0.49\textwidth]{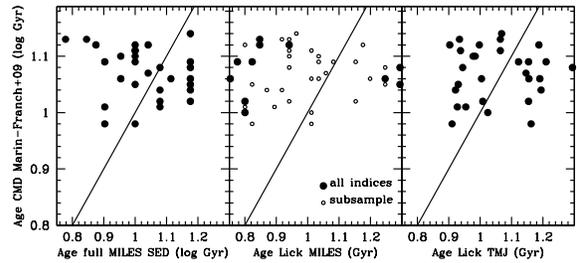}
  \caption{Same as Figure~\ref{fig:mastrolickagedeangeli} compared now with ages derived from CMD fit from Marin-Franch et al. (2009), where abundance ratio effects are explicitly considered.}
  \label{fig:mastrolickagemarin}
\end{figure}

\begin{figure}
  \includegraphics[width=0.49\textwidth]{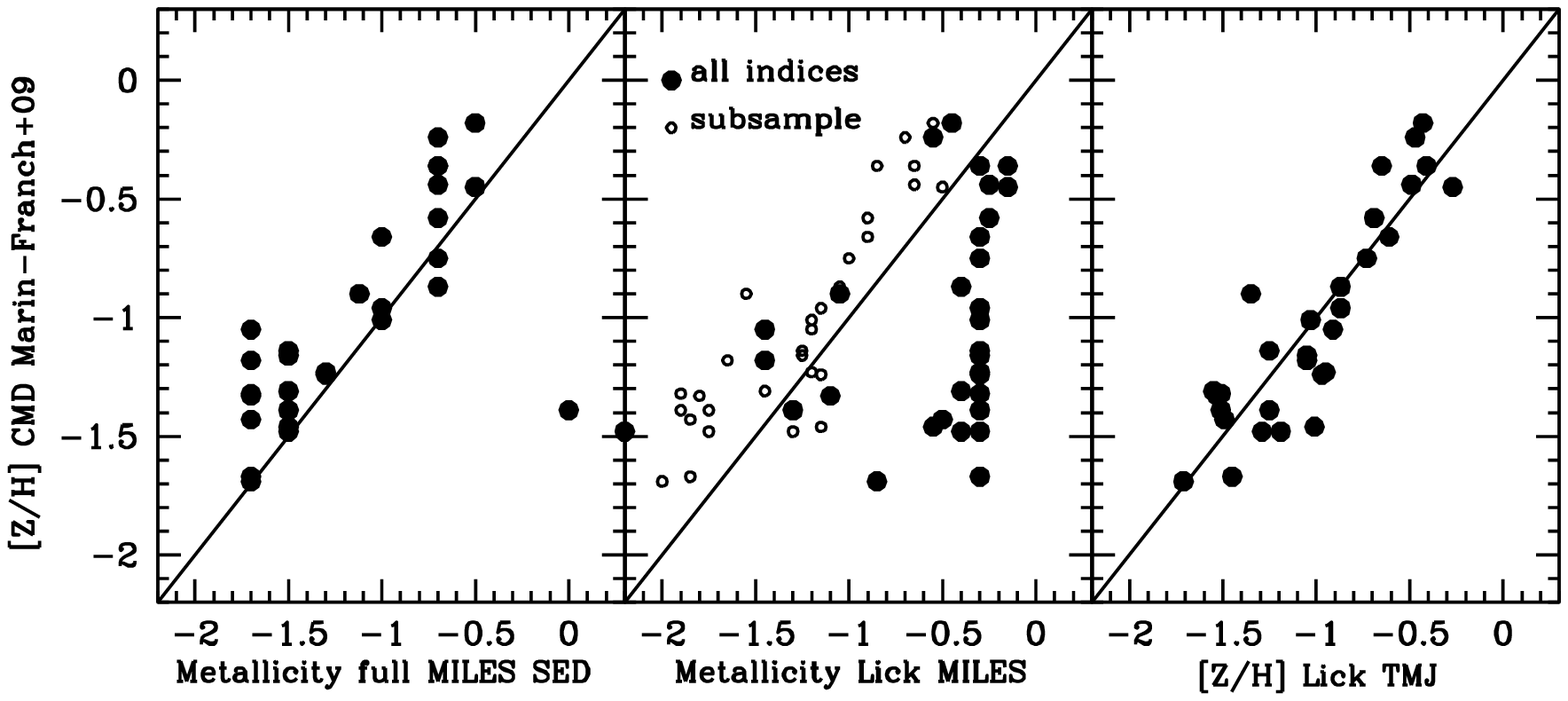}
  \caption{As in Figure~\ref{fig:mastrolickagemarin} for the metallicity published by Marin-Franch et al. (2009) which, after their correction for abundance-ratio effects is held to reflect the total metallicity [Z/H].}
  \label{fig:mastrolickmetalsmarin}
\end{figure}

What is interesting to evaluate here is - however - which ages and metallicities we get for MW GCs had we used - instead of the full SED spectral fitting - the Lick indices calculated on the same SED models (see Section~\ref{sec:lick}), i.e. without an explicit account of element abundance ratios. These are shown as filled symbols in the central panels of Figures~\ref{fig:mastrolickagedeangeli}, \ref{fig:mastrolickmetalsdeangeli}, \ref{fig:mastrolickagemarin}, 
\ref{fig:mastrolickmetalsmarin}. 
Ages are strongly underestimated with respect to the CMD ages ($y$-scale), which is due to the metallicity being overestimated (Figure~\ref{fig:mastrolickmetalsdeangeli}, central panel). The situation improves when considering a small subsample of indices (open symbols), the combination of which is little sensitive to element abundance ratios (namely, H$_{\delta_A}$, Mgb, Fe4383, Fe5335, Fe5406). This suitable index combination was defined by TJM11 as a base index set. 

Left-hand panels in the same figures shows the peak values of the marginalised age and metallicity $\chi^2$ distributions derived through full SED fit (cf. Figure 22). Right-hand panels show the values of age and total metallicity derived by \citet{tjm11} based on their model Lick absorption line indices. GCs ages, metallicities and [$\alpha/$Fe]-enhancement are derived through a $\chi^2$-fitting using the best-calibrated indices (see paper for details). 

Ages derived with both a full SED spectral fitting or using selected spectral lines with detailed modelling of abundance ratios are in much better agreement with the CMD-derived ones compared to those obtained using indices without a modelling of abundance ratios, even if the CMD-derived ages span a narrower range than those from integrated-light methods.\footnote{Note that TJM11 allowed the ages to run over 15 Gyr in extrapolation, while here we impose 15 Gyr as a maximum age. It is also interesting to note that the SED-fit is a bit sensitive to the adopted IMF, and the Kroupa one gives slightly smaller standard deviations, which makes sense. Standard deviations - after removal of the two outlying clusters NGC 5946 and 6544, are: age =  2.664; [Fe/H] =  0.226 for a Salpeter IMF; age =  2.51; [Fe/H] =  0.228 for a Kroupa IMF.} This gives us confidence in the intrinsic age-scale of the models.

Relevant to this test is to note that CMD-derived ages carry their own problems as there are zero-point issues depending on the adopted tracks and on whether element ratios are taken into account. Figure~\ref{fig:mastrolickagemarin} shows an age comparison identical to Figure~\ref{fig:mastrolickagedeangeli}, where we have used instead of the ages derived by De Angeli et al. (2005), those by Marin-Franch et al. (2009) in which a correction for abundance ratio effects has been included. Though the basic conclusions remain unchanged, the CMD-ages shift towards older values. More quantitatively, the comparison with the CMD ages of De Angeli et al. (2005) is better for the SED-fit case, where the medians of the difference (CMD-age $-$ integrated-age) are: $-0.03\pm$ 1.8 Gyr for the SED-fit and $-0.71\pm 2.83$~Gyr for the TMJ. They become comparable, however, if we use the latest age determination by Marin-Franch et al. (2009), with corresponding values of $1.5\pm 2.2$~Gyr $1.3\pm 2.76$~Gyr, respectively.

Far more striking is the comparison with the metallicity. Whereas in Figure~\ref{fig:mastrolickmetalsdeangeli} a remarkable agreement is evident between the Harris's (1996) metallicities and those derived with both the MILES-based SED-fitting and the Lick index fitting, when we adopt the metallicity scale of Marin-Franch et al. (2009) (Figure~\ref{fig:mastrolickmetalsmarin}) the SED-fit metallicities are systematically lower. This may highlight that the SED-fit-derived metallicities mostly reflect [Fe/H] rather than the total metallicity [Z/H] that both Marin-Franch et al. (2009) and the TMJ models aim to trace. This ambiguity in metallicity scales is amply discussed in \citet{CMetal03} and \citet{tjm11}. 

Finally, the values obtained with Lick indices directly calculated on the model SEDs without consideration of abundance ratio effects are off in any case (central panels in all figures).

These figures highlight four important facts, namely: i) when using GCs as test for SSP models - the CMD-derived quantities need to be taken with care; ii) the analysis of absorption line diagnostics should be carried out with models that account for element ratio effects otherwise the results are misleading; iii) selecting indices that - through models - are known not to be too sensitive to element ratio effects rectify the age and metallicity determination, which stresses again that the modelling of abundance ratio in the TMJ models is generally correct, and is required; iv) a good alternative appears to be the use of the full spectral fitting as the continuum shape helps getting ages and metallicity correctly. Methods iii) and iv) cannot however release the abundance ratio parameter. 

We have shown in Section~\ref{sec:lick} that the empirically-based models are solar-scaled at high metallicity. In order to check the models in this regime, we fit the composite spectrum of the old open cluster M67 \citep{ricardoetal04}, which is believed to have a metallicity about or just slightly below solar, solar-scaled element abundance ratios, and a CMD fitting age of 3 to 5 Gyr depending on the adopted isochrones and fitting details (see Chaboyer et al. 1999 for a review and Schiavon et al. 2004 for full references)\nocite{chaetal99}. 
\begin{figure*}
  \includegraphics[width=0.8\textwidth]{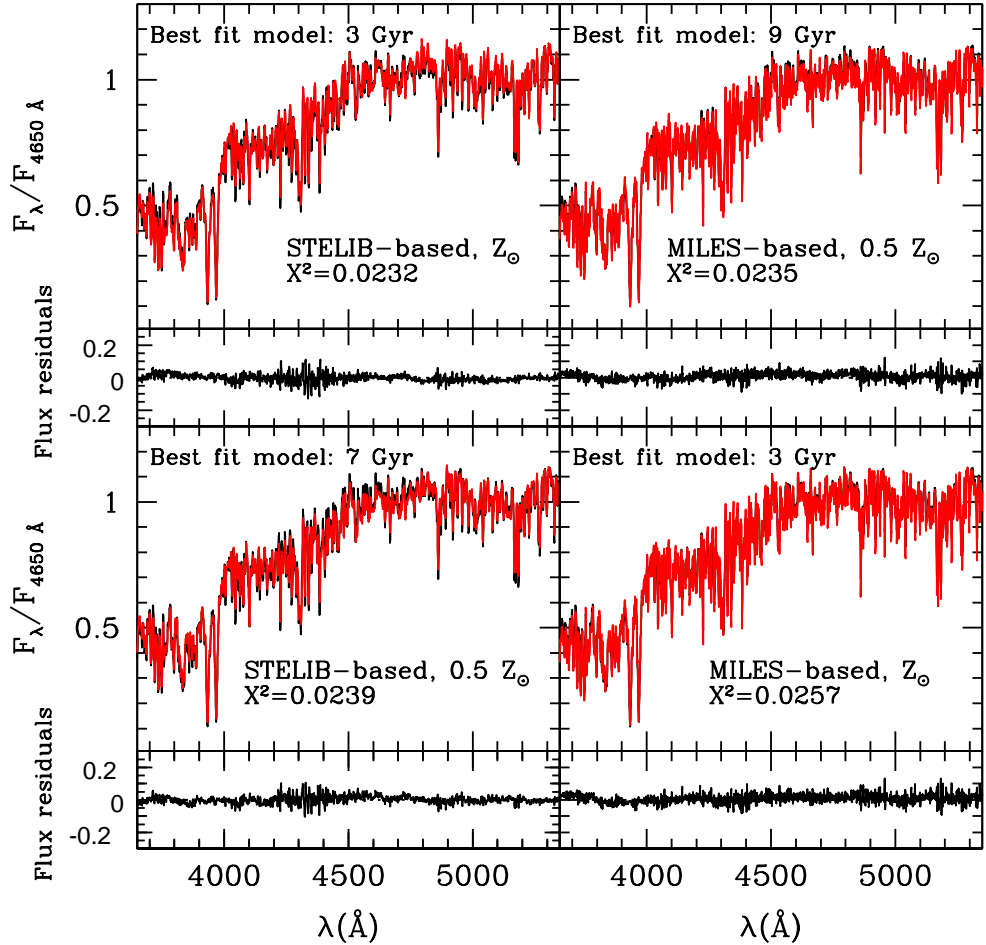}
  \caption{Model fits (red) to the composite spectrum of the solar metallicity old open cluster M67 (black, from Schiavon et al. 2004).}
  \label{fig:m67}
\end{figure*}
Figure~\ref{fig:m67} shows the best fit MILES-based and STELIB-based models for M67. The upper panels show the results for the minimum chi-square. STELIB-based models release a solution that is identical to the one from CMD-fitting. MILES-based models release a superb fit, which however gives the quite older age of 9 Gyr. The lower panels show the second ranked solution for STELIB-based and the 3rd ranked one for MILES-based models, which finally gets to the CMD solution. Note the risible difference in $\chi^2$.

\begin{figure}
  \includegraphics[width=0.49\textwidth]{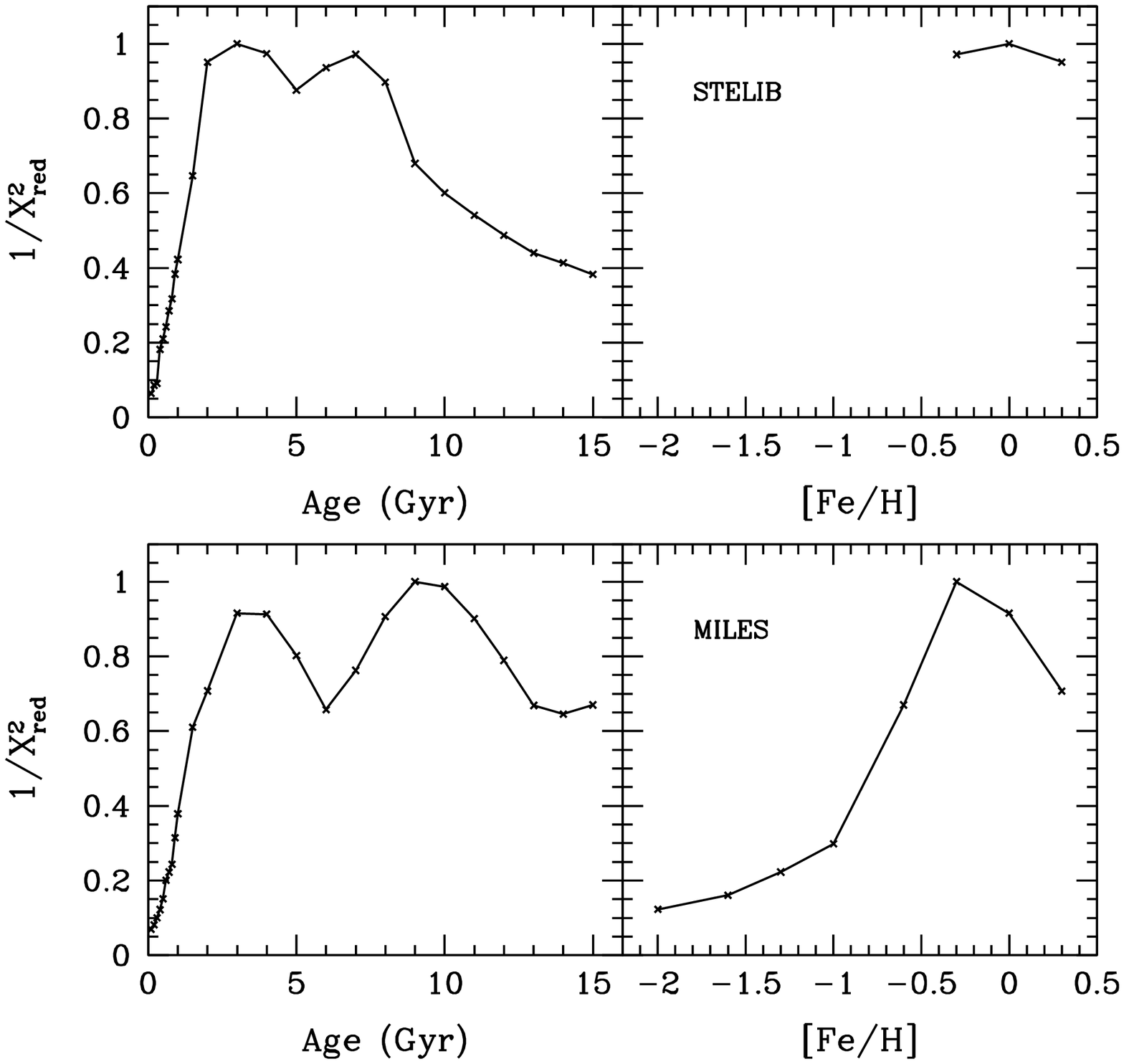}
  \caption{$\chi^{2}$ distribution of age and metallicity obtained for M67 with MILES-based (lower panels) and STELIB-based (upper panels) models.}
  \label{fig:errorm67}
\end{figure}

Figure~\ref{fig:errorm67} shows the $\chi^{2}$~distributions for age and metallicity as obtained with STELIB-based and MILES-based libraries. Obviously, taking into account a $68\%$~likelihood range instead of considering only the minimum 
$\chi^{2}$~would include the CMD solution for both type of models. We further verified that - using only solar metallicity MILES-based models - we obtain a best-fit age of 3 Gyr. Finally, we verified that using the STELIB models tied onto the MILES temperature scale still releases the CMD solution, meaning that the solution is not driven by the near-IR slope.

These results can be commented in several ways. One may drawn from here that the STELIB library better traces stars at solar metallicity, while MILES is a little too blue, which is why an older model is preferred. On the other hand, Vazdekis et al. (2010) claim to obtain the CMD solution as the best-fit. Here we note that the spectrum of M67 is not an observed integrated spectrum, but is a composite obtained by patching together individual stellar spectra via the population synthesis code of Schiavon et al. (2004). Differences between their code and ours, both in terms of method and adopted isochrones, could couple just with the MILES library in such a way as to explain why our first ranked solution is not consistent with theirs. The Vazdekis et al. (2010) model is similar to the Schiavon's one (i.e. isochrone synthesis on Padova isochrones). This event, though unlikely, is not excluded.

We refrain from drawing any conclusion on the libraries based on just one object, but we shall keep this fact in mind when fitting galaxies.

What Figure~\ref{fig:m67} shows is that all fits are excellent and hardly distinguishable, which illustrates two aspects of spectral fitting for age/metallicity determination, namely that: i) the sole consideration of the minimum $\chi^2$~is misleading (see Maraston et al. 2010 for a similar argument for the SED-fit of high-redshift galaxies)\nocite{CMetal10}; ii) the age-metallicity degeneracy is perfectly in place and the higher resolution of the models does not suffice to remove it. 

\subsubsection{Effect of different data on parameter derivation}
\begin{figure}
  \includegraphics[width=90mm]{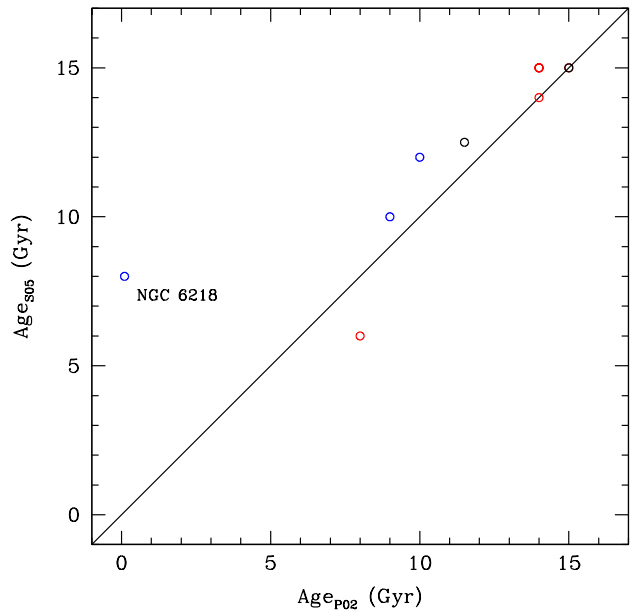}
  \caption{Comparison between ages derived for the cross-sample of 
    \citet{puetal02} and \citet{ricardoetal05} globular clusters using SED-fitting with 
    MILES-based models.}
  \label{fig:gcpuzschage}
\end{figure}

\begin{figure}
  \includegraphics[width=84mm]{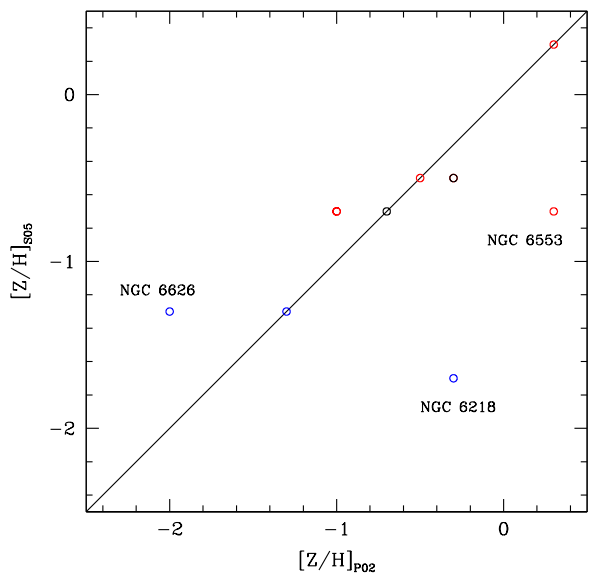}
  \caption{Same as Fig.\ref{fig:gcpuzschage}, but for metallicities.}
  \label{fig:gcpuzschfeh}
\end{figure}

The overlap between the \citet{puetal02} and the \citet{ricardoetal05} samples - consisting of eleven star clusters - allows 
a comparison between parameters derived using the same models but 
different observations. This is shown in Figures~\ref{fig:gcpuzschage} 
and~\ref{fig:gcpuzschfeh}. In most cases the outcome is the same, or at least 
a very similar, set of parameters. A few discrepancies can be seen, but 
considering that for these clusters there are noticeable differences in the 
shape of their spectra, and considering that even subtle differences can have 
an effect on the fitting, since age-metallicity degeneracies are large and 
$\chi^2$ distributions poorly constrained, the agreement is fairly  
good. The only exceptions are NGC~6218, whose SED shape in the P02 sample looks corrupted -- using the S05 spectrum we are able to recover parameters in good agreement with the literature values; the rebellious NGC~6553, for which the S05 version has been assigned a lower metallicity by $0.5$~dex, while the P02 version has been given a higher metallicity by the same amount; and NGC~6626, where once again the metallicity derived from the S05 spectrum is in better agreement with the \citet{har96} value.
\subsection{Globular cluster $B-V$ colours}
\label{gccolours}

We compare GC colours with the models to check whether the new spectra help in removing existing discrepancies. Notice that we limit ourselves to the $B-V$ colour as the models show a clear trend, whereas for the other colours there is either no change, or not a clear one (as e.g. for the $U-B$) or existing data are not sufficient (as for the near-$IR$ colours). 
\begin{figure*}
  \includegraphics[width=0.7\linewidth]{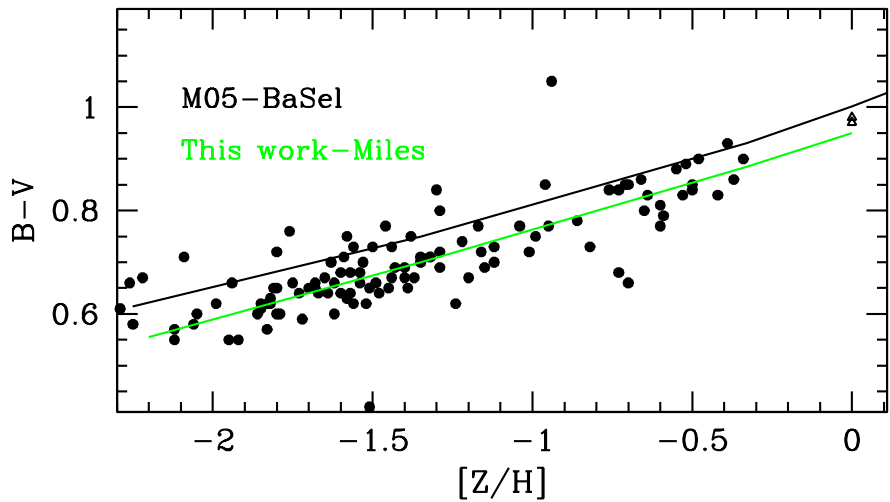}
  \caption{$B-V$ colours of M05 and MILES-based models as a function of metallicity for 12 Gyr of age (and a Salpeter IMF) compared to Milky Way GCs data \citep{har96}. This plot is identical to the correspondent panel of Figure~21 in M05. Note the improved agreement between models and GCs colours.}
  \label{fig:bvzh}
\end{figure*}
Figure~\ref{fig:bvzh} shows the $B-V$~colours of MW GCs as a function of metallicity, in a remake of the correspondent panel in Figure~21 of M05. Plotted are results for the MILES-based models (green) and for the M05 models (black). Consistently with the discussion in Section~5.3, the $B-V$ colours of the MILES-based models are bluer by 0.05 mag. and better fit the average locus of MW GCs. \citet{CM05} anticipated this result indicating a similar offset to be applied to the M05 models and it is neat to be able to find a reason for it.

Furthermore, \citet{peaetal11} show that the $g-r$~colours of M31 globular clusters are also much better matched by the new models because of the same effect, and \citet{sagetal10} use the colour corrections and find a better match to the population of the old M31 Bulge. These results strengthen our conclusions regarding the mismatch between the Sloan colours of LRG galaxies and population models \citep[][and Section~5.3]{CMetal09b}.

\section{Summary and Conclusions}
\label{summary}
Stellar population models find ubiquitous applications in astrophysics. The derivation of galaxy and star cluster properties such as ages, metallicities, star formation histories, stellar masses via decoding of the integrated energy emission is a crucial step in studies of galaxy formation and evolution and observational cosmology. Progress in technology now permits to obtain high spectral resolution for galaxies over a large range in masses and evolutionary status, i.e. age and level of on-going star formation, and for a relatively large range in redshift. Stellar population models have to be upgraded to meet the observational demand, and this was part of the aim for this work.

The models have three main ingredients stemming from stellar evolution, namely the stellar energetics, the stellar atmospheric parameters and the individual stellar spectra, and also depend on the algorithm adopted in the code and on a variety of other assumptions (e.g. the IMF, the HB morphology, the mass-loss parameters, etc., see Maraston 2005 for a wide discussion). The spectral resolution of the models is linked to the assumed individual stellar spectra. In this work we have varied the input stellar library - exploring a wide set of available empirical stellar libraries - keeping all other assumptions  as in the M05 modelling. 

We replenish missing stellar types in the libraries with theoretical spectra from the MARCS \citep{Gusetal08} theoretical library.
We additionally calculate stellar population models fully based on the MARCS library which can be used to extend the wavelength baseline of the empirically-based models to longer wavelengths, all along to the $K$-band and beyond. Similarly, we extend the models into the $UV$~by using our previously calculated high-resolution models based on the theoretical UVBLUE-Kurucz library \citep{CMetal09b}. 

This is not the first work of this kind. Bruzual \& Charlot (2003) presented the first set of higher resolution models exploiting the STELIB and the Pickles (1998) libraries, which have been the basis for many galaxy evolution studies. Vazdekis (1999) calculated models using the empirical stellar library by \citet{jon99}. \citet{CMetal09a} present very high-resolution SEDs up to $\sim~4500$ \AA\ using theoretical Kurucz-type spectra. Vazdekis et al. (2010) provide their models based on the MILES library.

The difference between this work and other similar in the literature is that - by exploring a wide range of empirical stellar libraries - we could also assess their effect on a stellar population model, which could be regarded as the second goal of this work. 

In general, the spectra of stars in common to the various libraries present discrepancies in selected spectral regions, which are shown and discussed, most of which get smoothed out in the integrated stellar population models, which display - at the end - a fair level of homogeneity. However, some differences remain, which will be interesting to test with data of galaxies and star clusters. For example, MILES-based population models display lower flux at their near-IR edge - long-ward of $\sim 6000 \AA$ - with respect to population models based on both the empirical STELIB or Pickles libraries as well as to those based on the theoretical library MARCS. We could trace this effect as mainly due to the temperature scale adopted for RGB stars in MILES.

As a relevant result, we confirm our earlier finding that the visual region of the spectrum of models based on empirical spectral libraries has lower flux (in a normalised sense) with respect to the BaSeL-Kurucz or the original Kurucz library \citep{CMetal09a}. The only exception are the models based on the ELODIE library, that however have such a high resolution that flux calibration becomes more difficult. This effect helped to a better match of the $g,r,i$~colours of Luminous Red Galaxies at redshift 0.4 and here we show that it also improves the comparison with the $B-V$ vs [Z/H] relation for Milky Way GCs. This conclusion is in agreement with the finding of Peacock et al. (2011) that the models of this paper better trace the $g,r$~colours of globular clusters in M31.

Most importantly, in this work we find that the theoretical library MARCS agrees well with the empirical libraries. This guarantees that the effect does not originate from the complicated procedure required to insert an empirical library in a population synthesis code. The adopted procedure matters, however. We have used the stellar parameters provided by the libraries and scaled the empirical spectra using the theoretical flux in a well-defined region. Alternative approaches match the empirical spectra to theoretical ones to insert the empirical libraries in the code (as done by Bruzual \& Charlot~2003, for example). In this way, differences between empirical and theoretical spectra are diluted.

We also show that other recent models based on MILES - namely those by Vazdekis et al. (2010) and by Conroy \& Gunn (2010) - confirm the \citet{CMetal09a} results, i.e. the impact of using empirical atmospheres on population models, even if details maybe different. Curiously, Conroy \& Gunn (2010) came to the opposite conclusion. 

We also discuss extensively the Lick indices calculated directly on the integrated MILES-based SEDs and compare them with those calculated using the fitting function (FF) approach as well as with element ratio sensitive index models (Thomas, Maraston \& Johansson 2011). In general, we find an excellent agreement between FF-based and SED-based index models. We also find a good agreement with the models by TMJ, if the metallicity dependent chemical pattern of the Milky Way stars is properly taken into account in this comparison. SED-based models agree with solar-scaled TMJ models at high-metallicity and with $[\alpha/Fe]-$enhanced models at sub-solar metallicity, because stellar libraries carry into the models the spectral pattern imprinted of the element abundance ratios of the Milky Way field stars they are built upon. 

We tested the age and metallicity scales of the models by fitting the observed SEDs of Milky Way GCs with ages and metallicities independently known from isochrone CMD fitting and resolved spectroscopy. We find overall good agreement when we perform full spectral fitting, suggesting that this an effective method to derive ages of stellar systems. This is because the spectral shape is age-sensitive even if details of spectral lines are not well-matched due to abundance ratio effects. Moreover, as we already noted -  these models are - qualitatively - [$\alpha$/Fe]-enhanced at sub-solar metallicities and solar-scaled at $Z_{\odot}$ and above, which also partly explains the good match to Milky Way GCs. For example, we could obtain excellent fits of the observed SED of the old open cluster M67 of the Milky Way disk, which has approximately a solar metallicity and 3 Gyr of age, and solar scaled abundance ratios. 

However, we find that - when we use - instead of the full spectral fitting - selected absorption lines without including a treatment of abundance-ratio effects - ages are underestimated and metallicities are overestimated. This is because absorption lines are very sensitive to element abundance ratio effects and peculiarities, which can be different between the stars composing the empirical libraries and external stellar systems such as star clusters and galaxies. A better agreement is achieved when one uses a sub-sample of indices the combination of which is little sensitive to element abundance ratios (namely H$_{\delta_A}$, Mgb, Fe4383, Fe5335, Fe5406, see Thomas, Johansson \& Maraston 2011 for details).
These facts should be considered when comparing these models to data. Using absorption line models including abundance ratio effects instead provide ages and metallicities in good agreement with the CMD-based ones (Thomas, Johansson, Maraston 2011) and the full spectral fitting performed with the SED models presented here. 

Finally, for M67 we noticed an interesting effect arising from the use of different libraries, namely that the models based on STELIB releases as best-fit (minimum $\chi^2$) a solution which is identical to the one from CMD fitting, whereas the MILES-based models give an older age and a lower metallicity, even if the solution with parameters as from the CMD fitting has a $\chi^2$ which is comparably good. This difference may originate from the MILES-based models being slightly bluer than the STELIB-based ones, hence releasing older ages. Obviously, a result based on just one object is merely a speculation, but it will be interesting to check whether such an effect persists using galaxy data, which we shall pursue in a future paper.

Models are available at www.maraston.eu/M11.
\section*{Acknowledgments}
During the model development and testing we could benefit from fruitful discussions with several colleagues, namely Daniel Thomas, Jonas Johansson, Issi Doyle, Jacopo Chevallard, Lucia Pozzetti, Harald Kuntschner, Lauren McArthur, Christy Tremonti, Rogerio Riffel, Andy Pickles, David Schlegel, Paul Crowther, Janine Pforr, Alessandra Beifiori, Ollie Steele, Rita Tojeiro, Will Percival, John Chisholm, Yanmei Chen, Guinevere Kauffmann. We thank Ricardo Schiavon for let us using their composite spectrum of M67. We are indebted to Bengt Edvardsson and the MARCS group for assisting us with their model library and providing us with additional computations and explanations. 
We thank an anonymous referee for a careful reading of the manuscript and for making valuable suggestions.

This work was supported by the Marie-Curie Excellence Team grant "UniMass", ref. MEXT-CT-2006-042754 of the Training and Mobility of Researchers programme financed by the European Community. 
%

\appendix

\section{Revising the near-IR slope of MILES-based models}\label{sec:corrmil}

Longward 6300~\AA, the MILES-based models brandish lower flux levels with respect to the other models and to the MARCS-based models (cf. Figure~\ref{fig:lrg}).
 
\begin{figure*}
  \centering
  \begin{minipage}{150mm}
    \includegraphics[width=140mm]{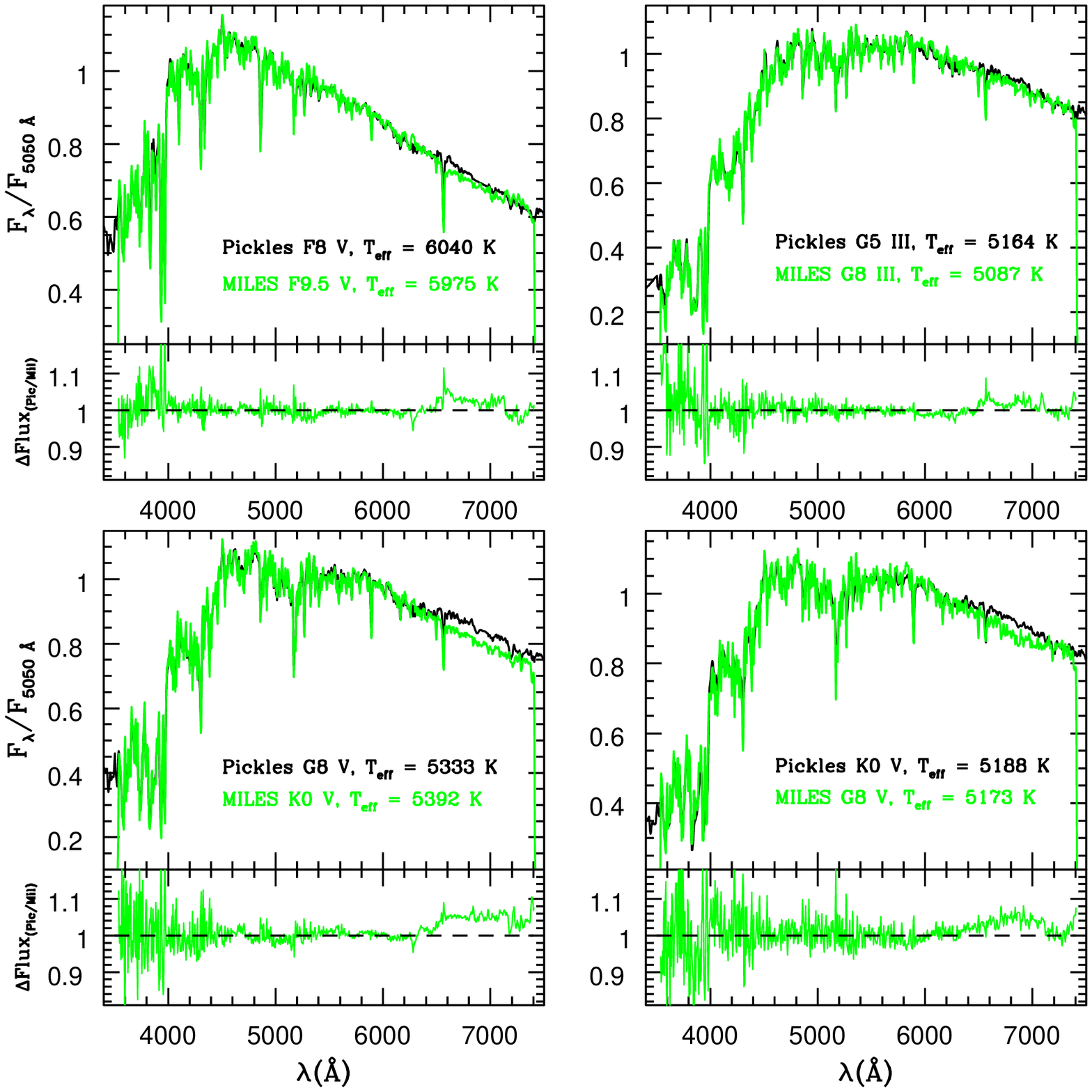}
    \caption{Best-fitting MILES spectra to four standard spectra from 
      \citet{pic98}, and their flux ratios as a function of wavelength, 
      defined as $\Delta F_{\lambda}=F_{\lambda,Pickles}/F_{\lambda,MILES}$ 
      ({\it smaller panels}). Flux ratios are scattered around~1, except 
      redwards of $\sim$6500 {\AA}, where a systematic flux deficit in the 
      MILES spectra can be observed. For this exercise the MILES spectra have 
      been smoothed and rebinned to the Pickles resolution and sampling.}
    \label{fig:mildip}
  \end{minipage}
\end{figure*}

\begin{figure}
  \includegraphics[width=84mm]{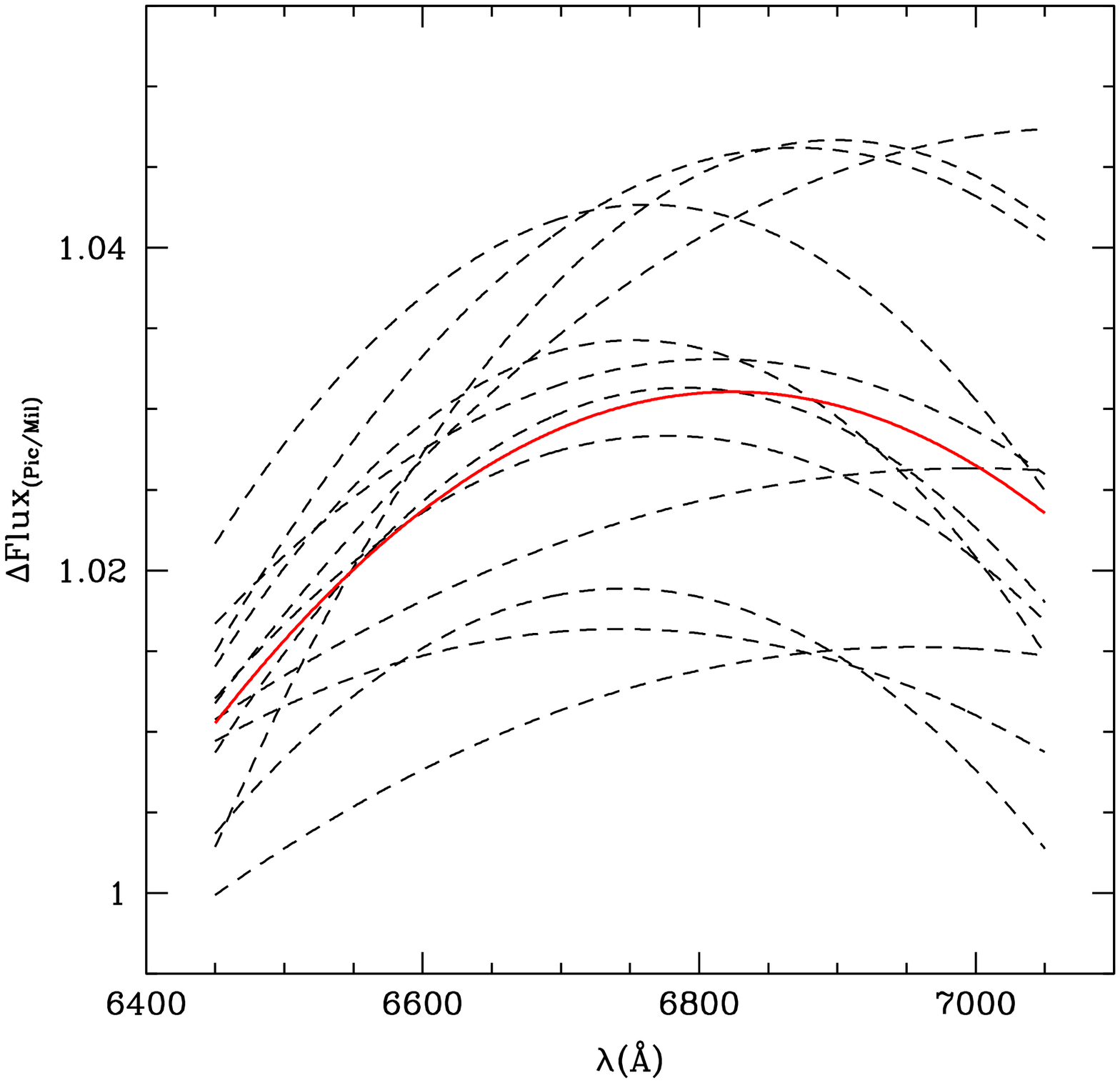}
  \caption{Second-order polynomial fits to the flux ratio plots of 
    best-fitting MILES spectra to 6 dwarf and 6 giant standard spectra of 
    \citet{pic98}, in the wavelength range 6450-7050 A ({\it dashed black 
    lines}). The straight average applied to all MILES stars is also shown 
    ({\it red solid line}).}
  \label{fig:ratiocurves}
\end{figure}

We traced this behaviour to have three origins:
\begin{enumerate}
\item {\it The temperature calibration}. For each \citet{pic98} standard 
  spectrum (we remind the reader that these are averages of several different 
  stars of the same spectral type and luminosity class, and as such, should be 
  less vulnerable to stochastic variations) we obtained the best-fitting 
  MILES spectrum in terms of minimising the sum of squared residuals. In 
  general, it appears that the MILES temperatures are slightly underestimated 
  with respect to the Pickles temperatures. In particular, this holds true for 
  Red Giant Branch bump and tip stars, whose impact on the 
  integrated near-IR light of old stellar populations is
  significant. A 
  spectral library with, on average, cooler temperatures, will have the 
  effect that the integrated SED becomes bluer. This connects nicely to the 
  discussion about the temperature scale in Sections~\ref{sec:fluxcal} and 3.6. Thus, by 
  replacing the MILES temperatures with the Pickles ones for these stars, we are able to 
  achieve higher flux levels red-ward of 6200 {\AA}.
\item {\it Gaps in stellar parameter space.} By shifting the coolest giants to 
  slightly higher temperatures, there are no longer any cool enough stars to 
  cover the parameter space required by the adopted stellar evolution 
  prescription for the tip of the RGB and E-AGB. By adding sufficiently cool 
  stars, for example from the Pickles library, part of the flux drop can be 
  cured.
\item {\it Short wavelength range systematic offset.} Besides the general lower flux long-ward 6300 \AA\, we spotted another near-IR feature on MILES spectra, namely a small dip in the flux levels between 6500 and 7000 {\AA}. This feature emerged when we compared  MILES spectra - smoothed to the appropriate resolution - with Pickles spectra. We found them to agree perfectly over the entire wavelength 
range, except in that short interval. This discrepancy 
can certainly not be ascribed to any difference in temperature, gravity, 
etc., but must be an intrinsic difference between Pickles and MILES spectra. 
Since the Pickles spectra are averages of several observations from numerous 
different sources, while the MILES spectra have all been obtained using the 
same instrumental setup, the most likely explanation is that this is an artefact 
originating from the MILES instrumental setup. Alternatively, this could be due to the adopted procedure for correcting against 
telluric absorption. Note that we spotted this mini-dip only on some MILES spectra as - in order to notice it - one needs to have an otherwise identical spectrum at moderate resolution, such as a Pickles spectrum. The dip is probably present in all MILES spectra - we argue it is a systematic - but we could not check all spectra in this respect for the above named reasons.

Fig. \ref{fig:mildip} shows such best-fitting MILES spectra to four standard 
\citet{pic98} spectra, accompanied by their respective flux ratios as a 
function of wavelength. MILES spectra have been smoothed and 
rebinned to the Pickles approximate resolution of R$\approx$500, and 5{\AA} 
sampling. It is obvious that, even when the two spectral energy 
distributions agree on general very well, there is a systematic deviation 
rewards $\sim$~6500 {\AA}. To counterbalance this, we have 
chosen 12 of the best-fitting MILES spectra (6 dwarfs and 6 giants), fitted 
second-order polynomials to the flux ratios in the aforementioned spectral 
range for each of the 12 components, and applied the straight average of 
these fitting curves (see Fig.\ref{fig:ratiocurves}) to all the MILES stars. 

\end{enumerate}

Based on the above corrections, for the solar metallicity case we provide two versions of our MILES-based models, original and revised. Figure A3 shows how the correspondent spectra compare, for the illustrative age of 12 Gyr (all other ages behave the same way). One notice from the residual plot that - since the most affected stars are cool giants - the near-IR portion of the spectrum is the only one that is affected appreciably by our revision. We recall that the same exercise at different metallicities cannot be made as it is based on Pickles spectra and they exist mostly at solar metallicity. It will be interesting to understand the physical meaning of such revision - if any -, which can be accomplished by comparing these models with stellar, star cluster and galaxy data.

\begin{figure}
    \includegraphics[width=84mm]{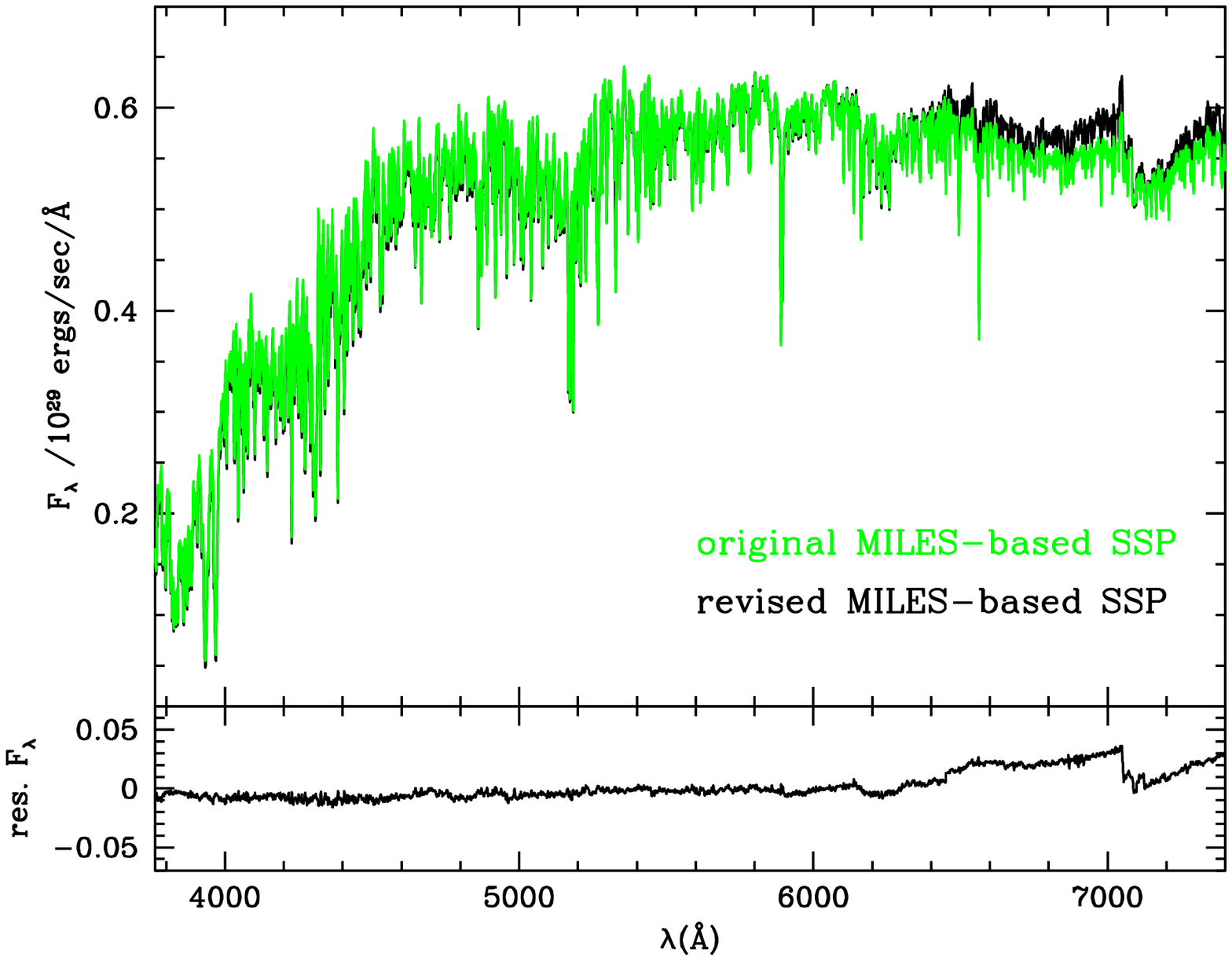}
    \caption{Comparison between MILES-based and MILES-revised based solar metallicity SSPs for the illustrative age of 12 Gyr The difference in flux (shown in the lower plot as $F_{\rm revised}-F$) induced by our revision is very mild all over the spectrum and mostly affect the near-IR part.}
    \label{fig:comprev}
\end{figure}

\bsp

\label{lastpage}

\end{document}